\documentclass[conference]{IEEEtran}
\usepackage{todonotes}
\usepackage{comment}
\usepackage{multirow}
\usepackage{hyperref}
\usepackage{textcomp}
\usepackage{proof,amssymb,stmaryrd,amsmath}
\usepackage{listings}
\usepackage{amsthm}
\usepackage{times}
\usepackage{caption,subcaption,graphicx}
\usepackage{frameit}
\usepackage{float}
\usepackage{pifont}

\usepackage[ruled,vlined,linesnumbered]{algorithm2e}

\title{\textsc{sGuard}: Towards Fixing Vulnerable Smart Contracts Automatically}

\author{
    \IEEEauthorblockN{Tai D. Nguyen, Long H. Pham, Jun Sun}
    \IEEEauthorblockA{dtnguyen.2019@smu.edu.sg, \{longph1989, sunjunhqq\}@gmail.com \\
        Singapore Management University, Singapore
    }
}

\newcommand{\tool}{\textsc{sGuard}}
\newcommand{\xmark}{\ding{55}}%

\SetCommentSty{mycommfont}
\theoremstyle{definition}
\newtheorem{definition}{Definition}
\newtheorem{lemma}{Lemma}
\newtheorem{theorem}{Theorem}
\newtheorem{example}{Example}[section]

\begin{document}

\maketitle

\begin{abstract}
Smart contracts are distributed, self-enforcing programs executing on top of blockchain networks. They have the potential to revolutionize many industries such as financial institutes and supply chains. However, smart contracts are subject to code-based vulnerabilities, which casts a shadow on its applications. As smart contracts are unpatchable (due to the immutability of blockchain), it is essential that smart contracts are guaranteed to be free of vulnerabilities. Unfortunately, smart contract languages such as Solidity are Turing-complete, which implies that verifying them statically is infeasible. Thus, alternative approaches must be developed to provide the guarantee. In this work, we develop an approach which automatically transforms smart contracts so that they are provably free of 4 common kinds of vulnerabilities. The key idea is to apply run-time verification in an efficient and provably correct manner. Experiment results with 5000 smart contracts show that our approach incurs minor run-time overhead in terms of time (i.e., 14.79\%) and gas (i.e., 0.79\%).    
 
\end{abstract}
	
\section{Introduction}

Blockchain is a public list of records which are linked together.
Thanks to the underlying cryptography mechanism, the records in the blockchain can resist against modification.
Ethereum is a platform which allows programmers to write distributed, self-enforcing programs (a.k.a smart contracts) executing on top of the blockchain network.
Smart contracts, once deployed on the blockchain network, become an unchangeable commitment between the involving parties.
Because of that, they have the potential to revolutionize many industries such as financial institutes and supply chains.
However, like traditional programs, smart contracts are subject to code-based vulnerabilities, which may cause huge financial loss and hinder its applications.
The problem is even worse considering that smart contracts are unpatchable once they are deployed on the network.
In other words, it is essential that smart contracts are guaranteed to be free of vulnerabilities before they are deployed.

In recent years, researchers have proposed multiple approaches to ensure smart contracts are vulnerability-free. These approaches can be roughly classified into two groups, i.e., verification and testing. However, existing efforts do not provide the required guarantee. Verification of smart contracts is often infeasible since smart contracts are written in Turing-complete programming languages (such as Solidity which is the most popular smart contract language), whereas it is known that testing (of smart contracts or otherwise) only shows the presence not the absence of vulnerabilities. 

In this work, we propose an approach and a tool, called \tool, which automatically fixes potentially vulnerable smart contracts. \tool{} is inspired by program fixing techniques for traditional programs such as C or Java, and yet are designed specifically for smart contracts. 
First, \tool{} is designed to guarantee the correctness of the fixes. Existing program fixing approaches (e.g., GenFrog~\cite{Weimer09icse}, PAR~\cite{ kim2013automatic}, Sapfix~\cite{marginean2019sapfix}) often suffer from the problem of weak specifications, i.e., a test suite is taken as the correctness specification. A fix driven by such a weak correctness criteria may over-fit the given test suites and does not provide correctness guarantee in all cases. Furthermore, fixes for smart contracts may suffer from not only time overhead but also gas overhead (i.e., extra fees for running the additional code) and \tool{} is designed to minimize the run-time overhead in terms of time and gas introduced by the fixes. 

Given a smart contract, at the high level, \tool{} works in two steps. In the first step, \tool~first collects a finite set of symbolic execution traces of the smart contract and then performs static analysis on the collected traces to identify potential vulnerabilities. 
As of now, \tool~supports 4 types of common vulnerabilities. Note that our static analysis engine is built from scratch as extending existing static analysis engines for smart contracts (e.g., Securify~\cite{tsankov2018securify} and Ethainter~\cite{brent2020ethainter}) for our purpose is infeasible. For instance, their sets of semantic rules are incomplete and sometimes produce conflicting results (i.e. a contract both complies and violates a security rule). In addition, they perform abstract interpretation locally (i.e., context/path-insensitive analysis) and thus suffer from many false positives. A contract fixed based on the analysis results from these tools may introduce unnecessary overhead.

In the second step, \tool~applies a specific fixing pattern for each type of vulnerability on the source code to guarantee that the smart contract is free of those vulnerabilities. Our approach is proved to be sound and complete on termination for the vulnerabilities that \tool~supports. 

To summarize, our contribution in this work is as follows.

\begin{itemize}
    \item We propose an approach to fix 4 types of vulnerabilities in smart contracts automatically.
    \item We prove that our approach is sound and complete for the considered vulnerabilities.
    \item We implement our approach as a self-contained tool, which is then evaluated with 5000 smart contracts. The experiment results show that \tool{} fixes 1605 smart contracts. Furthermore, the fixes incur minor run-time overhead in terms of time (i.e., 14.79\% on average) and gas (i.e., 0.79\%).
\end{itemize}

The remainder of the paper is organized as follows. In Section~\ref{overview}, we provide some background about smart contracts and illustrate how our approach works through examples. The problem is then defined formally in Section~\ref{definition}. In Section~\ref{details}, we present the details of our approach. The experiment results are presented in Section~\ref{results}. We discuss related work in Section~\ref{related} and conclude in Section~\ref{conclusion}.
\section{Background and Overview}
\label{overview}
\newcommand{\term}[1]{\textsc{#1}}
\newcommand{\conf}[1]{\langle #1 \rangle}
\newcommand{\eval}[1]{[ #1 ]}
\newcommand{\ctx}{\Gamma}
\newcommand{\nxt}{\leadsto}
\newcommand{\lst}{\Box}
\newcommand{\wrong}{\blacksquare}
\newcommand{\gas}{\mathcal{G}}
\newcommand{\defi}{\stackrel{\triangle}{=}}
\newcommand{\prim}{\mathsf{basic}}
\newcommand{\kil}{\times}
\newcommand{\hide}[1]{}
\newcommand{\env}{\mathcal{E}}
\newcommand{\bal}{\mathcal{B}}
\newcommand{\sife}[3]{{\bf{if}}~#1~{\bf{then}}~#2~{\bf{else}}~#3}
\newcommand{\swhile}[2]{{\bf{while}}~#1~{\bf{do}}~#2}
\newcommand{\fback}{{\bf{fallback}}}
In this section, we introduce relevant background on smart contracts and illustrate how our approach addresses the problem of smart contract vulnerabilities through examples. 

\subsection{Smart Contract}
The concept of smart contracts came into being with Ethereum~\cite{wood2014ethereum}, i.e., a digital currency platform with the capability of executing programmable code. It is subsequently supported by platforms such as RSK~\cite{rsk} and Hyperledger~\cite{hyperledger}. In this work, we focus on Ethereum smart contracts as it remains the most popular smart contracts platform. 

Intuitively speaking, an Ethereum smart contract implements a set of rules for managing digital assets in Ethereum accounts. In the Ethereum platform, there are two type of accounts, i.e., externally owned accounts (EOAs) and contract accounts. Both types of accounts have a 256-bit unique address and a balance which represents the amount of Ether (a.k.a. Ethereum currency unit) in the account. Contract accounts are the ones which are associated with smart contracts that can be used to perform certain predefined tasks. A smart contract is similar to a class in object-oriented programming languages such as Java or C\#. It contains persistent data such as storage variables and functions that can modify these variables (including a constructor which initializes them). Functions that are declared public can be invoked from other accounts (either EOAs or other contract accounts) through transactions, i.e., a sequence of function invocations. 

The Etherum platform supports multiple programming languages for smart contracts programming. Currently, the most popular one is Solidity, i.e., a Turing-complete programming language. For instance, Figure~\ref{fig:transfer_proxy}(a) shows a public function in a contract named \texttt{SmartMesh} written in Solidity. Once invoked, the function transfers certain amount of tokens from an account (at address \texttt{from}) to another account (at address \texttt{to}). A Solidity contract is compiled into Ethereum bytecode. With the bytecode, a transaction is then executed by the Ethereum Virtual Machine (EVM) on miners' machines. In its essence, EVM is a stack-based machine. Its details can be referred to in Section~\ref{semantics}. 

Solidity programs have a number of language features which are specific to smart contracts and are often associated with vulnerabilities. For instance, a public function marked with keyword $payable$ is allowed to receive Ether when it is invoked. The amount of Ether received is represented in the value of variable $msg.value$. That is, if an account invokes a $payable$ function of a contract and sets the value of $msg.value$ greater than 0, Ether is transferred from the invoking account to the invoked account.
Besides that, Ether can also be sent to other contracts using function $send()$ or $transfer()$ which are globally defined.
Note that in such a case, a specific no-name function, called the $fallback$ function, is executed if it is defined in the receiving contract. Note that a $fallback$ function is meant to be a safety valve when a non-existing function is called upon the contract, although it seems to be a source of problems instead. 
Furthermore, to prevent the harmful exploit of the network such as running infinite loops, each bytecode instruction (called opcode) is associated with a running cost called gas, which is paid from the caller's account.

\subsection{Vulnerabilities}
Just like traditional programs, smart contracts are subject to code-based vulnerabilities. A variety of vulnerabilities have been identified in real-world smart contracts, some of which have been exploited by attackers and have caused significant financial losses (e.g.,~\cite{deops199,contractbugs}). In the following, we introduce two kinds of vulnerabilities through examples. 



\definecolor{verylightgray}{rgb}{.97,.97,.97}

\lstdefinelanguage{Solidity}{
	keywords=[1]{anonymous, assembly, assert, balance, break, call, callcode, case, catch, class, constant, continue, constructor, contract, debugger, default, delegatecall, delete, do, else, emit, event, experimental, export, external, false, finally, for, function, gas, if, implements, import, in, indexed, instanceof, interface, internal, is, length, library, log0, log1, log2, log3, log4, memory, modifier, new, payable, pragma, private, protected, public, pure, push, require, return, returns, revert, selfdestruct, send, solidity, storage, struct, suicide, super, switch, then, this, throw, transfer, true, try, typeof, using, value, view, while, with, addmod, ecrecover, keccak256, mulmod, ripemd160, sha256, sha3}, 
	keywordstyle=[1]\color{blue}\bfseries,
	keywords=[2]{address, bool, byte, bytes, bytes1, bytes2, bytes3, bytes4, bytes5, bytes6, bytes7, bytes8, bytes9, bytes10, bytes11, bytes12, bytes13, bytes14, bytes15, bytes16, bytes17, bytes18, bytes19, bytes20, bytes21, bytes22, bytes23, bytes24, bytes25, bytes26, bytes27, bytes28, bytes29, bytes30, bytes31, bytes32, enum, int, int8, int16, int24, int32, int40, int48, int56, int64, int72, int80, int88, int96, int104, int112, int120, int128, int136, int144, int152, int160, int168, int176, int184, int192, int200, int208, int216, int224, int232, int240, int248, int256, mapping, string, uint, uint8, uint16, uint24, uint32, uint40, uint48, uint56, uint64, uint72, uint80, uint88, uint96, uint104, uint112, uint120, uint128, uint136, uint144, uint152, uint160, uint168, uint176, uint184, uint192, uint200, uint208, uint216, uint224, uint232, uint240, uint248, uint256, var, void, ether, finney, szabo, wei, days, hours, minutes, seconds, weeks, years},	
	keywordstyle=[2]\color{teal}\bfseries,
	keywords=[3]{block, blockhash, coinbase, difficulty, gaslimit, number, timestamp, msg, data, gas, sender, sig, value, now, tx, gasprice, origin},	
	keywordstyle=[3]\color{violet}\bfseries,
	identifierstyle=\color{black},
	sensitive=false,
	comment=[l]{//},
	morecomment=[s]{/*}{*/},
	commentstyle=\color{gray}\ttfamily,
	stringstyle=\color{red}\ttfamily,
	morestring=[b]',
	morestring=[b]"
}

\lstset{
	language=Solidity,
	backgroundcolor=\color{verylightgray},
	extendedchars=true,
	basicstyle=\scriptsize\ttfamily,
	showstringspaces=false,
	showspaces=false,
	numbers=left,
	numberstyle=\scriptsize,
	numbersep=9pt,
	tabsize=2,
	breaklines=true,
	showtabs=false,
	captionpos=b
}

\begin{figure*}[t]
    \parbox{.45\textwidth}{
\begin{lstlisting}
function transferProxy(address from, address to, uint value, uint fee) public {
  if (balances[from] < fee + value) revert();
  uint nonce = nonces[from];
  if (balances[to] + value < balances[to]) revert();
  if (balances[msg.sender] + fee < balances[msg.sender]) revert();
  balances[to] += value;
  balances[msg.sender] += fee;
  balances[from] -= value + fee;
  nonces[from] = nonce + 1;
}
\end{lstlisting}
    }
 ~~~~
    \raggedleft
    \parbox{.45\textwidth}{
\begin{lstlisting}
function transferProxy(address from, address to, uint value, uint fee) public {
  if (balances[from] < add_uint256(fee, value)) revert();
  uint nonce = nonces[from];
  if (add_uint256(balances[to], value) < balances[to]) revert();
  if (add_uint256(balances[msg.sender], fee) < balances[msg.sender]) revert();
  balances[to] = add_uint256(balances[to], value);
  balances[msg.sender] = add_uint256(balances[msg.sender], fee);
  balances[from] = sub_uint256(balances[from], add_uint256(_value, fee))
  nonces[from] = nonce + 1;
}
\end{lstlisting}
    }
    \\
    (a) Before ~~~~~~~~~~~~~~~~~~~~~~~~~~~~~~~~~~~~~~~~~~~~~~~~~~(b) After~~~~~~~~~~~~~~~~~~~~~~~~~~~~~~~~~~~~ \\
    \caption{CVE-2018-10376 patched by \textsc{sGuard}}
    \label{fig:transfer_proxy}
\end{figure*}


\begin{example}
One category of vulnerabilities is arithmetic vulnerability, e.g., overflow. For instance, in April 2018, an attacker exploited an integer overflow bug in a smart contract named \texttt{SmartMesh} and stole a massive amount of tokens (i.e., digital currency). The same bug affected 9 tradable tokens at that time and was named as \texttt{ProxyOverflow}. Figure~\ref{fig:transfer_proxy}(a) shows the (simplified) function \texttt{transferProxy} in the \texttt{SmartMesh} contract which contains the bug. The function is designed for transferring tokens from one account to another, while paying certain fee to the sender (see lines 6 and 7). The developer was apparently aware of potential overflow and introduced relevant checks at lines 2, 4 and 5. Unfortunately, one subtle bug is missed by the checks. That is, if \texttt{fee+value} is 0 (due to overflow) and \texttt{balances[from]=0}, the attacker is able to bypass the check at line 2 and subsequently increase the balance of \texttt{msg.sender} and \texttt{to} (see lines 6 and 7) by an amount more than \texttt{balances[from]}. During the attack, this bug was exploited to create tokens out of air. \emph{This example highlights that manually-written checks could be error-prone.} 
\end{example}

\begin{example}
Reentrancy vulnerability is arguably the most infamous vulnerability for smart contracts. It happens when a smart contract $C$ invokes a function of another contract $D$ and subsequently a call back (e.g., through the $fallback$ function in contract $D$) to contract $C$ is made while it is in an inconsistent state, e.g., the balance of contract $C$ is not updated. 
Figure~\ref{fig:mas_burner_fixed}(a) shows a part of a smart contract named \texttt{MasBurn} which contains a cross-function reentrancy vulnerability.
\texttt{MasBurn} implements a Midas protocol token, i.e., a tradable ERC20 token. It allows token holders to burn their owned tokens by sending tokens to a specific \texttt{BURN\_ADDRESS}, as shown at line 17. The total amount of burned tokens within one week can not exceed \texttt{weeklyLimit} (see line 16), which is a variable that limits the amount of tokens to be burned weekly. However, the problem is that the returned value of the function \texttt{getThisWeekBurnAmountLeft} (see line 16) has a data dependency on variable \texttt{numOfBurns}, and would be wrongly calculated in the case of a reentrancy call at line 17. That is, if the $fallback$ function of the contract at \texttt{BURN\_ADDRESS} contains a call back to the function \texttt{burn}, the function \texttt{getThisWeekBurnAmountLeft} is called with an outdated value of \texttt{numOfBurns}. As a result, the amount of burned tokens would exceed what is allowed. Although no Ether is lost (or created from air) in such an attack, the (implicit) specification of \texttt{MasBurn} is violated in such a scenario. This example also shows the difficulty in handling reentrancy vulnerability, i.e., whether a reentrancy is a vulnerability may depend on the specification of the contract. 
\end{example}

\subsection{Patching Smart Contracts} \label{sec:patchexample}
In the following, we illustrate how \textsc{sGuard} patches smart contracts through the two examples mentioned above. The technical details are presented in Section~\ref{details}. 
We remark that \textsc{sGuard} identifies vulnerabilities based on bytecode while patches them based on the corresponding source code. This is because analysis based on the bytecode is more precise than analysis based on the source code (as the former is not affected by bugs or optimizations in the Solidity compiler), whereas patching at the source code is transparent to the users. 

\begin{example}
The result of patching the function shown in Figure~\ref{fig:transfer_proxy}(a) using \textsc{sGuard} is shown in Figure~\ref{fig:transfer_proxy}(b). Almost all arithmetic operations (in statements or expressions) are replaced with function calls that perform the corresponding operations safely (i.e., with proper checks for arithmetic overflow or underflow). This effectively prevents the vulnerability as the function reverts immediately if \texttt{fee+value} overflows at line 2.
Note that the addition at line 9 is not patched as the variable \texttt{nonces} is not deemed critical itself or is depended on by some critical variables. 

One might argue that some of the modifications are not necessary, e.g., the one at line 4. This is true for this smart contract, if the goal is to prevent this particular vulnerability. In general, whether a modification is necessary or not can only be answered when the specification of the smart contract is present. \textsc{sGuard} does not require the specification from the user as that would limit its applicability in practice. \textsc{sGuard} thus always conservatively assumes all arithmetic overflow that may lead to vulnerability are problematic.  
Although this patch is not minimal, we guarantee that the patched \texttt{transferProxy} is free of arithmetic vulnerability. 
\end{example}

\begin{figure*}[t]
    \parbox{.45\textwidth}{
\begin{lstlisting}
function getThisWeekBurnedAmount() public view returns(uint) {
  uint thisWeekStartTime = getThisWeekStartTime();
  uint total = 0;
  for (uint i = numOfBurns; i >= 1; i--) {
    if (burnTimestampArr[i - 1] < thisWeekStartTime) break;
    total += burnAmountArr[i - 1];
  }
  return total;
}

function getThisWeekBurnAmountLeft() public view returns(uint) {
  return weeklyLimit - getThisWeekBurnedAmount();
}

function burn(uint amount) external payable {
  require(amount <= getThisWeekBurnAmountLeft());
  require(IERC20(tokenAddress).transferFrom(msg.sender, BURN_ADDRESS, amount));
  ++numOfBurns;
}
\end{lstlisting}
    }
    ~~~~
    \raggedleft
    \parbox{.45\textwidth}{
\begin{lstlisting}
function getThisWeekBurnedAmount() public view returns(uint) {
  uint thisWeekStartTime = getThisWeekStartTime();
  uint total = 0;
  for (uint i = numOfBurns; i >= 1; (i = sub_uint256(i, 1))) {
    if (burnTimestampArr[sub_uint256(i, 1)] < thisWeekStartTime) break;
    total = add_uint256(total, burnAmountArr[sub_uint256(i, 1)]);
  }
  return total;
}

function getThisWeekBurnAmountLeft() public view returns(uint) {
  return sub_uint256(weeklyLimit, getThisWeekBurnedAmount());
}

function burn(uint amount) external payable nonReentrant {
  require(amount <= getThisWeekBurnAmountLeft());
  require(IERC20(tokenAddress).transferFrom(msg.sender, BURN_ADDRESS, amount));
  (numOfBurns = add_uint256(numOfBurns, 1));
}
\end{lstlisting}
    } \\
    (a) Before ~~~~~~~~~~~~~~~~~~~~~~~~~~~~~~~~~~~~~~~~~~~~~~~~~~(b) After~~~~~~~~~~~~~~~~~~~~~~~~~~~~~~~~~~~~ \\
    \caption{MasBurn patched by \textsc{sGuard}}
    \label{fig:mas_burner_fixed}
\end{figure*}

\begin{example}
The result of applying \textsc{sGuard} to the contract shown in Figure~\ref{fig:mas_burner_fixed}(a)  is shown in Figure~\ref{fig:mas_burner_fixed}(b). \textsc{sGuard} identifies line 17 as an external call, which is critical as an external call invokes a function of another contract which might be under the control of an attacker. \textsc{sGuard} systematically identifies variables that the external call at line 17 depends on (either through control dependency or data dependency). Afterwards, \textsc{sGuard} patches these variables and operations accordingly. 
In particular, this external call has control dependency on the if-statement at line 5 and is followed by a storage update (\texttt{++numOfBurns} at line 18). 
\begin{itemize}
    \item The subtractions at lines 4, 5, 6, 12 are replaced with calls of function \texttt{sub\_uint256}, which checks underflow.
    \item The additions at lines 6, 18 are replaced with calls of function \texttt{add\_uint256} to avoid overflow.
    \item Function \texttt{burn} is patched to prevent reentrancy. That is, we introduce the modifier \texttt{nonReentrant} at line 15. This modifier is derived from OpenZeppelin \cite{openzeppelin}, a library for secure smart contract development.
\end{itemize}
The resultant smart contract is free of arithmetic vulnerability and reentrancy vulnerability.
\end{example}

\section{Problem Definition}
\label{definition}
In the following, we first present the semantics for Solidity smart contracts, and then define our problem. 

\subsection{Concrete Semantics} \label{semantics}
A smart contract ${\cal S}$ can be viewed as a finite state machine ${\cal S} = (Var, init, N, i, E)$ where $Var$ is a set of variables; $init$ is the initial valuation of the variables; $N$ is a finite set of control locations; $i \in N$ is the initial control location, i.e., the start of the contract; and $E \subseteq N \times C \times N$ is a set of labeled edges, each of which is of the form $(n, c, n')$ where $c$ is an opcode. 
There are a total of 78 opcodes in Solidity (as of version 0.5.3), as summarized in Table~\ref{tbl:op}. Note that each opcode is statically assigned with a unique program counter, i.e., each opcode can be uniquely identified based on the program counter. 

Note that $Var$ includes stack variables, memory variables, and storage variables. Stack variables are mostly used to store primitive values and memory variables are used to store array-like values (declared explicitly with keyword $memory$). Both stack and memory variables are volatile, i.e., they are cleared after each transaction. In contrast, storage variables are non-volatile, i.e., they are persistent on the blockchain. Together, the variables' values identify the state of the smart contract at a specific point of time. At the Solidity source code level, stack and memory variables can be considered as local variables in a specific function; and storage variables can be considered as contract-level variables. 

A concrete trace of the smart contract is an alternating sequence of states and opcodes $\langle s_0, op_0, s_1, op_1, \cdots \rangle$ such that each state $s_i$ is of the form $(pc_i,S_i,M_i,R_i)$ where $pc_i \in N$ is the program counter; $S_i$ is the valuation of the stack variables; $M_i$ is the valuation of the memory variables; and $R_i$ is the valuation of the storage variables. Note that the initial state $s_0$ is $(0,S_0,M_0,R_0)$ where $S_0$, $M_0$ and $R_0$ are the initial valuation of the variables defined by $init$. Furthermore, for all $i$, $(pc_{i+1},S_{i+1},M_{i+1},R_{i+1})$ is the result of executing opcode $op_i$ given the state $(pc_i,S_i,M_i,R_i)$ according to the semantic of $op_i$. The semantics of opcodes are shown in Figure~\ref{fig:op} in form of execution rules, each of which is associated with a specific opcode. Each rule is composed of multiple conditions above the line and a state change below the line. The state change is read from left to right, i.e., the state on the left changes to the state on the right if the conditions above the line are satisfied. Note that this formal semantics is based on the recent effort on formalizing Etherum~\cite{9152785}. 

Most of the rules are self-explanatory and thus we skip the details and refer the readers to~\cite{9152785}. It is worth mentioning how external calls are abstracted in our semantic model. Given an external function call (i.e., opcode \term{CALL}), the execution temporarily switches to an execution of the invoked contract. The result of the external call, abstracted as $res$, is pushed to the stack. 




\begin{figure*}[h]
\begin{frameit}
\small\centering
$$
\infer[\term{STOP}]
{
(pc,S,M,R) \nxt \lst
}
{
\begin{array}{c}
\end{array}
}
\quad
\infer[\term{POP}]
{
(pc,S,M,R) \nxt (pc+1,S_1,M,R)
}
{
\begin{array}{c}
S_1,x = S.pop()
\end{array}
}
\quad
\infer[\term{UNARY-op}]
{
(pc,S,M,R) \nxt (pc+1,S_2,M,R)
}
{
\begin{array}{c}
S_1,x = S.pop()~~z = op(x)~~S_2 = S_1.push(z)
\end{array}
}
$$
$$
\infer[\term{BINARY-op}]
{
(pc,S,M,R,pc) \nxt (pc+1,S_3,M,R,pc+1)
}
{
\begin{array}{c}
S_1,x = S.pop()~~~~S_2,y = S_1.pop()\\z = op(x,y)~~~~S_3 = S_2.push(z)
\end{array}
}
\quad
\infer[\term{TERNARY-op}]
{
(pc,S,M,R) \nxt (pc+1,S_4,M,R)
}
{
\begin{array}{c}
S_1,x = S.pop()~~S_2,y = S_1.pop()~~S_3,m = S_2.pop()\\z = op(x,y,m)~~~~S_4 = S_3.push(z)
\end{array}
}
$$
$$
\infer[\term{MLOAD}]
{
(pc,S,M,R) \nxt (pc+1,S_2,M,R)
}
{
\begin{array}{c}
S_1,p = S.pop()~~v = M[p]~~S_2 = S_1.push(v)
\end{array}
}
\quad
\infer[\term{MSTORE}]
{
(pc,S,M,R) \nxt (pc+1,S_2,M_1,R)
}
{
\begin{array}{c}
S_1,p = S.pop()~~S_2,v = S_1.pop()~~M_1 = M[p \leftarrow v]
\end{array}
}
$$
$$
\infer[\term{SLOAD}]
{
(pc,S,M,R) \nxt (pc+1,S_2,M,R)
}
{
\begin{array}{c}
S_1,p = S.pop()~~v = R[p]~~S_2 = S_1.push(v)
\end{array}
}
\quad
\infer[\term{SSTORE}]
{
(pc,S,M,R) \nxt (pc+1,S_2,M,R_1)
}
{
\begin{array}{c}
S_1,p = S.pop()~~S_2,v = S_1.pop()~~R_1 = R[p \leftarrow v]
\end{array}
}
$$
$$
\infer[\term{DUP-i}]
{
(pc,S,M,R) \nxt (pc+1,S_1,M,R)
}
{
\begin{array}{c}
v = S.get(i)~~S_1 = S.push(v)
\end{array}
}
~~
\infer[\term{SWAP-i}]
{
(pc,S,M,R) \nxt (pc+1,S_2,M,R)
}
{
\begin{array}{c}
v_0 = S.get(0)~~v_i = S.get(i)~~S_1 = S[0 \leftarrow v_i]~~S_2 = S_1[i \leftarrow v_0]
\end{array}
}
$$
$$
\infer[\term{JUMPI-T}]
{
(pc,S,M,R) \nxt (lbl,S_2,M,R)
}
{
\begin{array}{c}
S_1,lbl = S.pop()~~S_2,c = S_1.pop()~~c \neq 0
\end{array}
}
~~
\infer[\term{JUMPI-F}]
{
(pc,S,M,R) \nxt (pc+1,S_2,M,R)
}
{
\begin{array}{c}
S_1,lbl = S.pop()~~S_2,c = S_1.pop()~~c = 0
\end{array}
}
$$
$$
\infer[\term{JUMP}]
{
(pc,S,M,R) \nxt (lbl,S_1,M,R)
}
{
\begin{array}{c}
S_1,lbl = S.pop()
\end{array}
}
\quad\quad\quad
\infer[\term{CALL}]
{
(pc,S,M,R) \nxt (pc+1,S_1,M,R)
}
{
\begin{array}{c}
res = call()~~S_1 = S.push(res)
\end{array}
}
$$
$$
\infer[\term{SHA3}]
{
(pc,S,M,R) \nxt (pc+1,S_3,M,R)
}
{
\begin{array}{c}
S_1,p = S.pop()~~S_2,n = S_1.pop()~~v = sha3(M[p,p+n])~~S_3 = S_2.push(v)
\end{array}
}
$$
\end{frameit}
\caption{Operational semantics of Ethereum opcodes. $pop$, $push$, and $get$ are self-explanatory stack operations. $m[p \leftarrow v]$ denote an operations which returns the same stack/mapping as $m$ except that the value of position/key $p$ is changed to $v$.
Rule \term{UNARY-op} (\term{BINARY-op}, \term{TERNARY-op}) applies to all unary (binary, ternary) operations; rule \term{DUP-i}, applies to all duplicate operations; and rule \term{SWAP-i} applies to all swap operations. 
}\label{fig:op}
\end{figure*}

\begin{table}[t]
{\scriptsize
\begin{tabular}{|p{2.1cm}|p{5.8cm}|}
\hline
Rule & Opcodes \\
\hline
STOP & \texttt{SELFDESTRUCT}, \texttt{REVERT}, \texttt{INVALID}, \texttt{RETURN}, \texttt{STOP}  \\ \hline
POP & \texttt{POP} \\ \hline
UNARY-op & \texttt{NOT}, \texttt{ISZERO}, \texttt{CALLDATALOAD}, \texttt{EXTCODESIZE}, \texttt{BLOCKHASH}, \texttt{BALANCE}, \texttt{EXTCODEHASH} \\ \hline
BINARY-op & \texttt{ADD}, \texttt{MUL}, \texttt{SUB}, \texttt{DIV}, \texttt{SDIV}, \texttt{MOD}, \texttt{SMOD}, \texttt{EXP}, \texttt{SIGNEXTEND}, \texttt{LT}, \texttt{GT}, \texttt{SLT}, \texttt{SGT}, \texttt{EQ}, \texttt{AND}, \texttt{OR}, \texttt{XOR}, \texttt{BYTE}, \texttt{SHL}, \texttt{SHR}, \texttt{SAR} \\ \hline
TERNARY-op & \texttt{ADDMOD}, \texttt{MULMOD}, \texttt{CALLDATACOPY}, \texttt{CODECOPY}, \texttt{RETURNDATACOPY} \\ \hline
MLOAD & \texttt{MLOAD} \\ \hline
SHA3 & \texttt{SHA3} \\ \hline
MSTORE & \texttt{MSTORE}, \texttt{MSTORE8} \\ \hline
SLOAD & \texttt{SLOAD} \\ \hline
SSTORE & \texttt{SSTORE} \\ \hline
DUP-I & \texttt{DUP1}$\cdots$\texttt{DUP16}\\ \hline
SWAP-I & \texttt{SWAP1}$\cdots$\texttt{SWAP16} \\ \hline
JUMPI-T/JUMPI-F & \texttt{JUMPI}\\ \hline
JUMP & \texttt{JUMP}\\ \hline
CALL & \texttt{STATICCALL}, \texttt{CALL}, \texttt{CALLCODE}, \texttt{CREATE}, \texttt{CREATE2}, \texttt{DELEGATECALL}, \texttt{SELFDESTRUCT} \\ \hline
\end{tabular}
}
\caption{The opcodes according to each rule}
\label{tbl:op}
\end{table}

\subsection{Symbolic Semantics}
\label{symbolic}
In order to define our problem, we must define the kinds of vulnerabilities that we focus on. Intuitively, we say that a smart contract suffers from certain vulnerability if there exists an execution of the smart contract that satisfies certain constraints. In the following, we extend the concrete traces to define symbolic traces of a smart contract so that we can define whether a symbolic trace suffers from certain vulnerability. 

To define symbolic traces, we first extend the concrete values to symbolic values.
Formally, a symbolic value has the form of $op(operand_0, \cdots, operand_n)$ where $op$ is an opcode and $operand_0, \cdots, operand_n$ are the operands. Each operand may be a concrete value (e.g., an integer number or an address) or a symbolic value. Note that if all operands of an opcode are concrete values, the symbolic value is a concrete value as well, i.e., the result of applying $op$ to the concrete operands. For instance, \texttt{ADD(5,6)} is $11$. Otherwise, the value is symbolic. 
One exception is that if $op$ is \texttt{MLOAD} or \texttt{SLOAD}, the result is symbolic even if the operands are concrete, as it is not trivial to maintain the concrete content of the memory or storage.
For instance, loading a value from address \texttt{0x00} from the storage results in the symbolic value \texttt{SLOAD(0x00)} and increasing the value at storage address \texttt{0x00} by 6 results in a symbolic value \texttt{ADD(SLOAD(0x00),0x06)}. For another instance, the result of symbolically executing \texttt{SHA3(n,p)} is \texttt{SHA3(MLOAD(n,p))}, i.e., the SHA3 hash of the value located from address $n$ to $n + p$ in the memory. 


With the above,
a symbolic trace is an alternating sequence of states and opcodes $\langle s_0, op_0, s_1, op_1, \cdots \rangle$ such that each state $s_i$ is of the form $(pc_i,S_i^s,M_i^s,R_i^s)$ where $pc_i$ is the program counter; $S_i^s$, $M_i^s$ and $R_i^s$ are the valuations of stack, memory and storage respectively. Note that $S_i^s$, $M_i^s$ and $R_i^s$ may hold symbolic values as well as concrete ones. 
For all $i$, $(pc_{i+1},S_{i+1}^s,M_{i+1}^s,R_{i+1}^s)$ is the result of executing opcode $op_i$ symbolically given the state $(pc_i,S_i^s,M_i^s,R_i^s)$. 

A symbolic execution engine is one which systematically generate the symbolic traces of a smart contract. Note that different from concrete execution, a symbolic execution would generate two traces given an \texttt{if}-statement, one visits the then-branch and the other visits the else-branch. Furthermore, in the case of an external call (i.e., \texttt{CALL}), instead of switching the current execution context to another smart contract, we can simply use a symbolic value to represent the returned value of the external call.

\subsection{Problem Definition} \label{sec:probdef}
Intuitively, a vulnerability occurs when there are dependencies from certain critical instructions (e.g., \texttt{CALL} and \texttt{DELEGATECALL}) to a set of specific instructions (e.g., \texttt{ADD}, \texttt{SUB} and \texttt{SSTORE}). Therefore, to define our problem, we first define (control and data) dependency, based on which we define the vulnerabilities. 
\begin{figure}[t]
    \centering
    \includegraphics[width=0.49\textwidth]{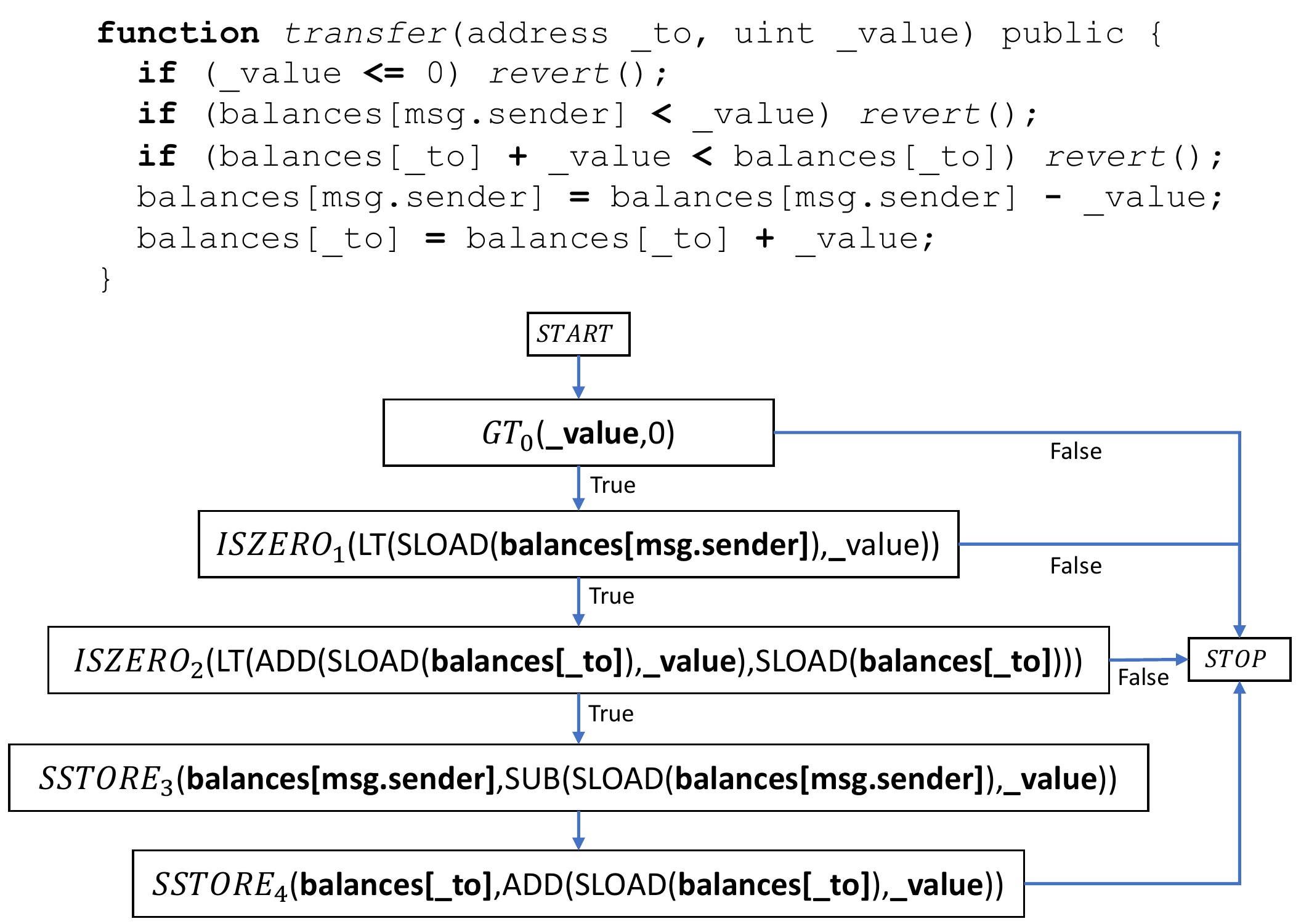}
    \caption{An example of control and data dependency}
    \label{fig:control_dependency}
\end{figure}

\begin{definition}[Control dependency]
\label{def:control_dependency}
An opcode $op_j$ is said to be control-dependent on $op_i$ if there exists an execution from $op_i$ to $op_j$ such that $op_j$ post-dominates all $op_k$ in the path from $op_i$ to $op_j$ (excluding $op_i$) but does not post-dominates $op_i$. An opcode $op_j$ is said to post-dominate an opcode $op_i$ if all traces starting from $op_i$ must go through $op_j$.
\end{definition}
Figure~\ref{fig:control_dependency} illustrates an example of control dependency. The source code is shown on the top and the corresponding control flow graph is shown on the bottom. All variables and their symbolic values are summarized in Table~\ref{tab:var_list}. The source code presents secure steps to transfer \texttt{\_value} tokens from \texttt{msg.sender} account to \texttt{\_to} account. There are 3 then-branches followed by 2 storage updates. According to the definition, both \texttt{SSTORE$_3$} and \texttt{SSTORE$_4$} are control-dependent on \texttt{ISZERO$_1$}, \texttt{ISZERO$_2$} and \texttt{GT$_0$}.

\begin{table}[t]
    \centering
    \begin{tabular}{|l|l|}
        \hline
        \textbf{Variable} & \textbf{Symbolic Value}  \\ \hline
        $\_to$ & \texttt{CALLDATALOAD(0x04)} \\ \hline
        $\_value$ & \texttt{CALLDATALOAD(0x24)} \\ \hline
        $balances[msg.sender]$ & \texttt{SHA3(MLOAD(0x00,0x40))} \\ \hline
        $balances[\_to]$ & \texttt{SHA3(MLOAD(0x00,0x40))} \\ \hline
    \end{tabular}
    \caption{Variables and their symbolic values of Figure~\ref{fig:control_dependency}}
    \label{tab:var_list}
\end{table}

\begin{definition}[Data dependency]
An opcode $op_j$ is said to be data-dependent on $op_i$ if there exists a trace which executes $op_i$ and subsequently $op_j$ such that $W(op_i) \cap R(op_j) \neq \emptyset$ where $R(op_j)$ is a set of locations read by $op_j$; $W(op_i)$ is a set of locations written by $op_i$.
\end{definition}

Figure~\ref{fig:control_dependency} also illustrates an example of data dependency. Opcode \texttt{ISZERO$_1$} and \texttt{ISZERO$_2$} are data-dependent on \texttt{SSTORE$_3$} and \texttt{SSTORE$_4$}. It has 2 traces, i.e., one trace loads data from storage address \texttt{SHA3(MLOAD(0x00,0x40))} which is  written by \texttt{SSTORE$_1$} and \texttt{SSTORE$_2$} in another trace.

We say an opcode $op_j$ is dependent on opcode $op_i$ if $op_j$ is control or data dependent on $op_i$ or $op_j$ is dependent on an opcode $op_k$ such that $op_k$ is dependent on $op_i$. \\

\noindent \emph{Vulnerabilities} In the following, we define the 4 kinds of vulnerabilities that we focus on, i.e., intra-function and cross-function reentrancy, dangerous tx.origin and arithmetic overflow. We remark that while we can certainly detect more kinds of vulnerabilities, it is not always clear how to fix them, i.e., it may not be feasible to know the intended behavior. For example, in the case of fixing an \textit{accessible selfdestruct} vulnerability (i.e., a smart contract suffers from this vulnerability if it may be destructed by anyone~\cite{brent2020ethainter}), we would not know for sure who should have the privilege to access \texttt{selfdestruct}. 

Let $C$ be a set of critical opcodes which contains \texttt{CALL}, \texttt{CALLCODE}, \texttt{DELEGATECALL}, \texttt{SELFDESTRUCT}, \texttt{CREATE} and \texttt{CREATE2}, i.e., the set of all opcode associated with external calls except \texttt{STATICCALL}. The reason that \texttt{STATICCALL} is excluded from $C$ is that \texttt{STATICCALL} can not update storage variables of the called smart contract and thus is considered to be safe.

\begin{figure}[t]
    \raggedleft
    \parbox{.45\textwidth}{
\begin{lstlisting}
uint numWithdraw = 0;
function withdraw() external {
    uint256 amount = balances[msg.sender];
    balances[msg.sender] = 0;
    (bool ret, ) = msg.sender.call.value(amount)("");
    require(ret);
    numWithdraw ++;
}
\end{lstlisting}
    }
    \caption{A non-reentrant case captured by NW}
    \label{fig:def_same_function_reentrancy}
\end{figure}

\begin{definition}[Intra-function reentrancy vulnerability]
\label{def:same_function_reentrancy}
A symbolic trace suffers from intra-function reentrancy vulnerability if it executes an opcode $op_c \in C$ and subsequently executes an opcode $op_s$ in the same function such that $op_s$ is $\texttt{SSTORE}$, and $op_c$ depends on $op_s$. 
\end{definition}
A smart contract suffers from intra-function reentrancy vulnerability if and only if at least one of its symbolic traces suffers from intra-function reentrancy vulnerability.
The above definition is inspired from the \emph{no writes after call} (NW) property~\cite{tsankov2018securify}. It is however more accurate than NW, as it avoids violations of NW which are not considered as reentrancy vulnerability. For instance, the function shown in Figure~\ref{fig:def_same_function_reentrancy} violates NW, although it is not subject to reentrancy vulnerability. It is because the external call \texttt{msg.sender.call} has no dependency on \texttt{numWithdraw}. In other words, there does not exist a dependency from $op_c$ to $op_s$.

\begin{figure}[t]
    \raggedleft
    \parbox{.45\textwidth}{
\begin{lstlisting}
function transfer(address to, uint amount) external {
    if (balances[msg.sender] >= amount) {
      balances[to] += amount;
      balances[msg.sender] -= amount;
    }
}

function withdraw() external nonReentrant {
    uint256 amount = balances[msg.sender];
    (bool ret, ) = msg.sender.call.value(amount)("");
    require(ret);
    balances[msg.sender] = 0;
}
\end{lstlisting}
    }
    \caption{An example of cross-function reentrancy vulnerability}
    \label{fig:def_cross_function_reentrancy}
\end{figure}

\begin{definition}[Cross-function reentrancy vulnerability]
\label{def:cross_function_reentrancy}
A symbolic trace $tr$ suffers from cross-function reentrancy vulnerability if it executes an opcode $op_s$ where $op_s$ is \texttt{SSTORE} and there exists a symbolic trace $tr'$ subject to intra-function reentrancy vulnerability such that the opcode $op_c$ of $tr'$ depends on $op_s$, and they belong to different functions.
\end{definition}
A smart contract suffers from cross-function reentrancy vulnerability if and only if at least one of its symbolic traces suffers from cross-function reentrancy vulnerability. This vulnerability differs from intra-function reentrancy as the attacker launches an attack through two different functions, which makes it harder to detect. Figure~\ref{fig:def_cross_function_reentrancy} shows an example of cross-function reentrancy. The developer is apparently aware of intra-function reentrancy and thus add the modifier \texttt{nonReentrant} to the function \texttt{withdraw} for preventing reentrancy. However, reentrancy is still possible through function \texttt{transfer}, in which case the attacker is able to double his Ether. That is, the attacker receives Ether at line 10 and illegally transfers it to another account at line 3. Although cross-function reentrancy vulnerabilities were described in Sereum~\cite{rodler2018sereum} and Consensys~\cite{known-attacks}, our work is the first work to define  it formally. 

\begin{figure}[t]
    \raggedleft
    \parbox{.45\textwidth}{
\begin{lstlisting}
function sendTo(address receiver, uint amount) public {
    require(tx.origin == owner);
    receiver.transfer(amount);
}
\end{lstlisting}
    }
    \caption{An example of dangerous tx.origin vulnerability}
    \label{fig:def_dangerous_tx_origin}
\end{figure}

\begin{definition}[Dangerous tx.origin vulnerability]
A symbolic trace suffers from dangerous tx.origin vulnerability if it executes an opcode $op_c \in C$ which depends on an opcode \texttt{ORIGIN}. 
\end{definition}
A smart contract suffers from dangerous tx.origin vulnerability if and only if at least one of its symbolic traces suffer from dangerous tx.origin vulnerability. This vulnerability happens due to an incorrect usage of the global variable \texttt{tx.origin} to authorize an user. An attack happens when a user $U$ sends a transaction to a malicious contract $A$, which intentionally forwards this transaction to a contract $B$ that relies on a vulnerable authorization check (e.g., \texttt{require(tx.origin == owner)}). Since \texttt{tx.origin} returns the address of $U$, contract $A$ successfully impersonates $U$. Figure~\ref{fig:def_dangerous_tx_origin} presents an example suffering from dangerous tx.orgin vulnerability, i.e., a malicious contract may impersonate the owner to withdraw all Ethers.

\begin{definition}[Arithmetic vulnerability]
\label{def:arithmetic}
A symbolic trace suffers from arithmetic vulnerability if it executes an opcode $op_c$ in $C$ and $op_c$ depends on an opcode $op_a$ which is \texttt{ADD}, \texttt{SUB}, \texttt{MUL} or \texttt{DIV}.
\end{definition}
A smart contract suffers from arithmetic vulnerability if and only if at least one of its symbolic traces suffer from arithmetic vulnerability. Intuitively, this vulnerability occurs when an external call data-depends on an arithmetic operation (e.g., addition, subtraction, or multiplication). For instance, the example in the Figure~\ref{fig:mas_burner_fixed} is vulnerable due to the presence of data dependency between the external call at line 17 and the expression \texttt{weeklyLimit - getThisWeekBurnedAmount()} at line 12. Arithmetic vulnerabilities are the target of multiple tools designed for vulnerability detection. In general, arithmetic vulnerability detection using static analysis often results in high false positive. Therefore, tools such as Securify~\cite{tsankov2018securify} and Ethainter~\cite{brent2020ethainter} do not support this vulnerability in spite of its importance. In the above definition, we focus on only critical arithmetic operations to reduce false positives. That is, an arithmetic operation is not considered critical as long as the smart contract does not spread its wrong computation to other smart contracts through external calls. For instance, wrong ERC20 token transfer (e.g., CVE-2018-10376) is not critical because it can be reverted by the contract's admin, whereas wrong Ether transfer is irreversible. \\


\noindent \emph{Problem definition} Our problem is then defined as follows. Given a smart contract $S$, construct a smart contract $T$ such that $T$ satisfies the following.
\begin{itemize}
    \item Soundness: $T$ is free of any of the above vulnerabilities. 
    \item Preciseness: For every symbolic trace $tr$ of $S$, if $tr$ does not suffer from any of the vulnerabilities, there exists a symbolic trace $tr'$ in $T$ which, given the same inputs, produces the same outputs and states.
    \item Efficiency: $T$'s execution speed and gas consumption are minimally different from those of $S$.
\end{itemize}
Note that the first two are about the correctness of construction, whereas the last one is about the performance in terms of computation and gas overhead.

\section{Detailed Approach}
\label{details}
In this section, we present the details of our approach. The key challenge is to precisely identify where vulnerabilities might arise and fix them accordingly. Note that precisely identifying control/data-dependency is a prerequisite for precisely identifying vulnerabilities.  
One approach to identify vulnerabilities is through static analysis based on over-approximation. For instance, multiple existing tools (e.g., Securify~\cite{tsankov2018securify} and Ethainter~\cite{brent2020ethainter}) over-approximate Etherum semantics using rewriting rules and leverage rewriting systems such as Datalog to identify vulnerabilities through critical pattern matching. While useful (and typically efficient) in detecting vulnerabilities, such approaches are not ideal for our purpose for multiple reasons. First, there are often many false alarms as they perform abstract interpretation locally (i.e., context/path-insensitive analysis). In our setting, once a vulnerability is identified, we fix it by introducing additional run-time checks. False alarms thus translate to runtime overhead in terms of both time and gas. Second, existing approaches are often incomplete, i.e., not all dependencies are captured. For instance, Securify ignores data dependency through storage variables, i.e., the dependency due to \texttt{SSTORE(c,b)} is lost if $c$ is not a constant, whereas Ethainter ignores control dependency completely. Thirdly, rewriting systems such as Datalog may terminate without any result, in which case the analysis result may not be sound. Therefore, in our work, we propose an algorithm which covers all dependencies with high precision and always terminates with the correct result.

The details of our algorithm is shown in Algorithm~\ref{algo:sguard}. From a high-level point of view, it works as follows. First, symbolic traces are systematically enumerated, up to certain threshold number of iterations for each loop. Second, each symbolic trace is checked to see whether it is subject to certain vulnerability according to our definitions. Lastly, the corresponding source code of the vulnerability is identified based on the AST and fixed. In the following, we present details of each step one-by-one.
\begin{algorithm}[t]
\small
    establish a bound for each loop\;
    enumerate symbolic traces $Tr$\;
    \ForEach{trace $tr$ in $Tr$}{
        let $dp \gets dependency(tr)$\;
        $fixReentrancy(tr, dp)$\;
        $fixTxOriginAndArithemic(tr, dp)$\;
    }
\caption{$sGuard$}
\label{algo:sguard}
\end{algorithm}

\subsection{Enumerating Symbolic Traces}
\label{sect:enumerate_symbolic_traces}
Note that our definitions of vulnerabilities are based on symbolic traces. Thus, in this first step, we set out to collect a set of symbolic traces $Tr$. As defined in Section~\ref{symbolic}, a symbolic trace is a sequence of the form $\langle s_0, op_0,\cdots, s_n, op_n, s_{n+1} \rangle$. In the following, we focus on symbolic traces that are maximum, i.e., the last opcode $op_n$ is either \texttt{REVERT}, \texttt{INVALID}, \texttt{SELFDESTRUCT}, \texttt{RETURN}, or \texttt{STOP}. 

Systematically generating the maximum symbolic traces is straightforward in the absence of loops, i.e., we simply apply the symbolic semantic rules iteratively until it terminates. 
In the presence of loops, however, as the condition to exit the loop is often symbolic, this procedure would not terminate. This is a well-known problem for symbolic execution and the remedy is typically to bound the number of iterations heuristically. Such an approach however does not work in our setting, since we must identify all data/control dependency to identify all potential vulnerabilities.  
In the following, we establish a bound on the number of iterations on the loops which we prove is sufficient for identifying the vulnerabilities that we focus on.

Given a smart contract ${\cal S} = (Var, init, N, i, E)$, a loop is in general a strongly connected component in ${\cal S}$. Thanks to structural programming, we can always identify the loop heads, i.e., the control location where a \emph{while}-loop starts or a recursive function is invoked. In the following, we associate each location $n \in N$ with a bound, denoted as $bound(n)$. If $n$ is a loop head, $bound(n)$ intuitively means how many times $n$ has to be visited in at least one of symbolic traces we collect. If $n$ is not part of any strongly connected component, we have $bound(n) = 1$. Otherwise, $bound(n)$ is defined as follows.
\begin{itemize}
    \item If $(n, op_n, n') \in E$ and $n'$ is the loop head, $bound(n) = 0$ if $op_n$ is not an assignment; otherwise $bound(n) = 1$.
    \item If $(n, op_n, n') \in E$, $n'$ is not the loop head and there is no $m$ such that $(n, op_n, m) \in E$, i.e., $n$ is not branching, $bound(n) = bound(n')$ if $op_n$ is not an assignment; otherwise $bound(n) = bound(n') + 1$.
    \item If $(n, op_n, m_0) \in E$ and $(n, op_n, m_1) \in E$, i.e., $n$ is branching, $bound(n) = bound(m_1) + bound(m_2)$.
\end{itemize}
Intuitively, the bound of a loop head is computed based on the number of branching statements and assignment statements inside the loop. That is, the bound of a loop head $n$ can be computed by traversing the CFG in the reverse order, i.e., from the exiting nodes of the loop to $n$. Every execution path maintains a bound, which equals to the number of assignment statements in that path. If two execution paths meet at a branching statement then the new bound is set to the sum of their bounds. In our implementation, the bounds for every node $n \in N$ are statically computed using a fixed-point algorithm, with a complexity of $\mathcal{O}((\#N)^2)$ where $\#N$ is the number of nodes. Once the bounds are computed, we systematically enumerate all maximum symbolic traces such that each loop head $n$ is visited at most $bound(n)$ times. It is straightforward to see that this procedure always terminates and returns a finite set of symbolic traces.  

\begin{example}
\begin{figure}[t]
    \centering
    \includegraphics[width=0.45\textwidth]{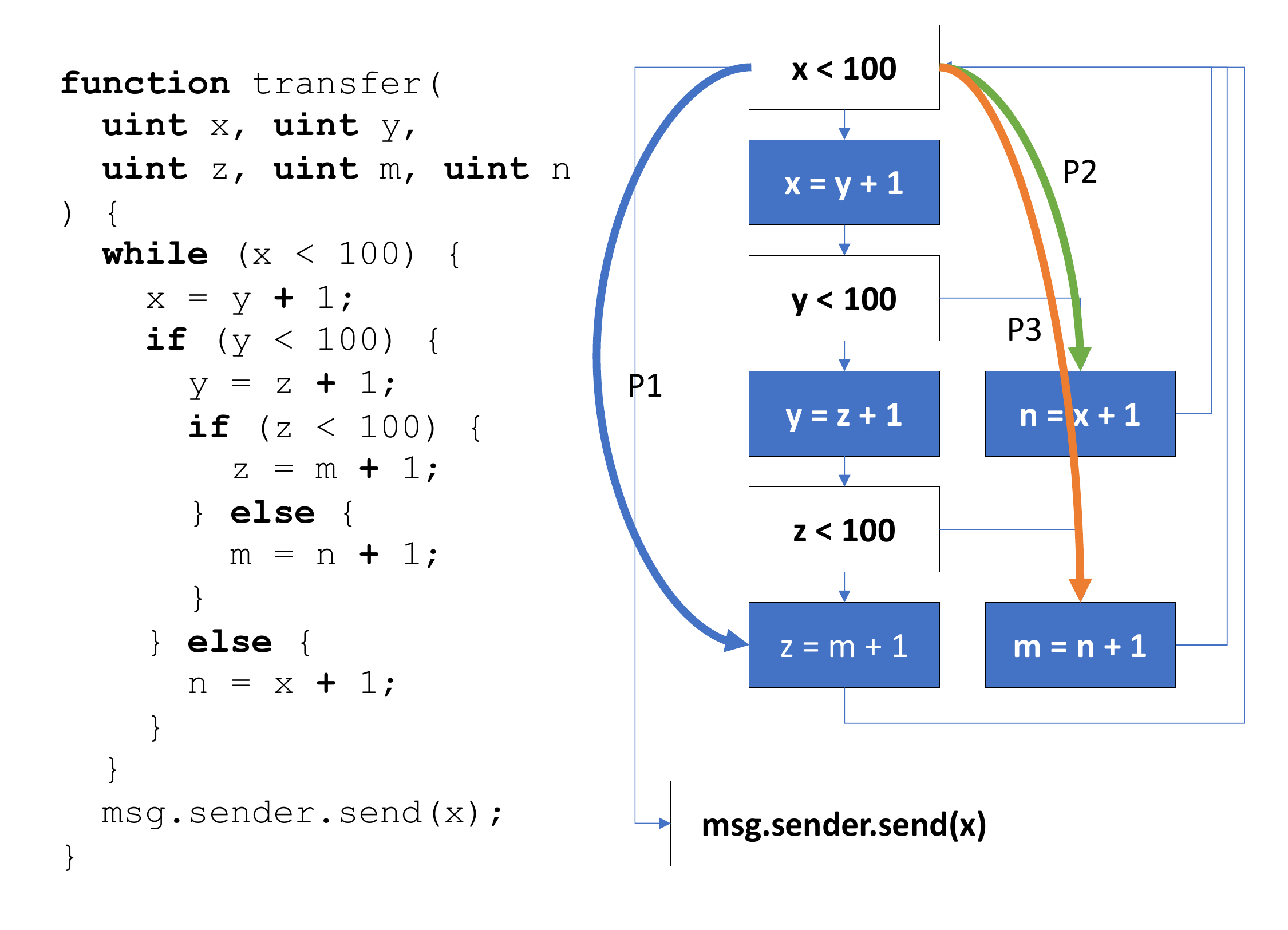}
    \caption{An example on how the $bound(n)$ is computed}
    \label{fig:boundary}
\end{figure}
In the following, we illustrate how $bound(x < 100)$ is computed. The example is shown in the Figure~\ref{fig:boundary} where the graph on the right represents the source code on the left (a.k.a.~control flow graph which can be constructed using existing approaches~\cite{nguyen2020sfuzz}). Assignment statements are highlighted in blue. There is a total of 3 paths $P1, P2, P3$ in the \emph{while-loop}, and they visit 5 assignment statements. Since we follow both branches of an if-statement, there exists a symbolic trace $tr$ containing $P1, P2, P3$ regardless of the order. Trace $tr$ is of the form $\langle \cdots, op_i, \cdots, op_j, \cdots, op_k, \cdots, op_i', \cdots \rangle$ where $op_i$ and $op_i'$ are executed opcodes of the loop head $x < 100$; $op_j$ is mapped to $y < 100$ and $op_k$ is mapped to $z < 100$.

There are 5 assignment statements between $op_i$ and $op_i'$ and the bound of the loop head is 5. Note that the number of assignment statements in the example is the number of \texttt{SWAP}s appeared in between $op_i$ and $op_i'$. 

\end{example}

The following establishes the soundness of our approach, i.e., using the bounds, we are guaranteed to never miss any of the 4 kinds of vulnerabilities that we focus on. 

\begin{lemma}
\label{lem:Tr_is_complete}
Given a smart contract, if there exists a symbolic trace which suffers from intra-function reentrancy vulnerability (or cross-function reentrancy, or dangerous tx.origin, or arithmetic vulnerability), there must be one in $Tr$.  \hfill \qedsymbol
\end{lemma}

We sketch the proof in the following. All vulnerabilities in Section~\ref{definition} are defined based on control/data dependency between opcodes. That means we always have a vulnerable trace, if there is, one as long as the set of symbolic traces we collect exhibit all possible dependency between opcodes. To see that all dependencies are exhibited in the traces we collect, we distinguish two cases. All control dependency between opcodes are identified as long as all possible branches in the smart contract are executed. This condition is satisfied based on the way we collect traces in $Tr$. This argument applies to data dependency between opcodes which do not belong to any loop as well. Next, we consider the data dependency between opcodes inside a loop. Note that with each loop iteration, there are two possible cases: no new data dependency is identified (i.e., the data dependency reaches fixed point) or at least 1 new dependency is identified. If the loop contains $n$ assignments, in the worst case, all of these opcodes depend on each other and we need a trace with $n$ iterations to identify all of them. Based on how we compute the bound for the loop heads, the trace is guaranteed to be in $Tr$. Thus, we establish that the above lemma is proved. 

It is well-known that symbolic execution engines may suffer from the path explosion problem. \tool~is not immune as well, i.e., the number of symbolic paths explored by \tool~is in general exponential in the loop bounds. Existing symbolic execution engines address the problem by allowing users to configure a bound $K$ which is the maximum number of times any loop is unrolled. In practice, it is highly non-trivial to know what $K$ value should be used. Given the impact of $K$, i.e., the number of paths are exponential in the value of $K$, existing tools often set $K$ to be a small number by default, such as 3 in sCompile~\cite{chang2019scompile} and 5 in Manticore~\cite{mossberg2019manticore}; and it is unlikely that users would configure it differently. While having a large $K$ leads to the path explosion problem, having a small $K$ leads to false negatives. For instance, with $K = 3$, the overflow vulnerabilities due to the two expressions $m = n + 1$, $n = x + 1$ in the Figure~\ref{fig:boundary} would be missed as this bound is not sufficient to infer dependency from variable $x$ on $m$ and $n$. In contrast, \tool{}~automatically identifies a loop bound for each loop which guarantees that no vulnerabilities are missed. In Section~\ref{results}, we empirically evaluate whether the path explosion problem occurs often in practice.

\subsection{Dependency Analysis}
Given the set of symbolic traces $Tr$, we then identify dependency between all opcodes in every symbolic trace in $Tr$, with the aim to check whether the trace suffers from any vulnerability. 
In the following, we present our approach to capture dependency from symbolic traces. 

Given a symbolic trace $Tr$, 
an opcode $op_i$, we aim to identify a set of opcodes $dp$ in $Tr$ such that: (soundness) for all $op_k$ in $dp$, $op_i$ depends on $op_k$; and
(completeness) for all $op_k$ in $Tr$, if $op_i$ depends on $op_k$ then $op_k \in dp$. To identify $dp$, we systematically identify all opcodes that $op_i$ is control-dependent on in $Tr$, all opcodes that $op_i$ is data-dependent on in $Tr$ and then compute their transitive closure. 

\begin{algorithm}[t]
\small
let $edges$ $\gets \emptyset$\; 
\ForEach{trace tr in Tr}{
    \ForEach{$op_i, pc_i$ in tr}{
        \If{$op_i$ = \texttt{JUMPI}} {
            let $edge$ $\gets (pc_i, pc_{i+1})$ \;
            add $edge$ to $edges$\;
        }
    }
}
return $edges$\;
\caption{build CFG}
\label{algo:build_cfg}
\end{algorithm}
To systematically identify all control-dependency, we build a control flow graph (CFG) from $Tr$ (as shown in Algorithm~\ref{algo:build_cfg}). Afterwards, we build a post-dominator tree based on the CFG using a workList algorithm~\cite{dominator}. The result is a set $PD(op_i)$ which are the opcodes that post-dominate $op_i$. The set of opcodes which $op_i$ control-depend on in the symbolic trace $tr$ is then systematically identified as the following. 
\begin{multline*}
\{~op~|~op \in tr; \exists~(op_m, op_n) \in succs(op), \\op_i \in PD(op_m), op_i \notin PD(op_n)~\}
\end{multline*}
where $succs(op)$ returns successors of $op$ according to CFG. 
\begin{algorithm}[t]
\small
let $opcodes \gets \emptyset$\;
\ForEach{$op_j$ that taints $op_i$} {
    \If{$op_j$ is an assignment opcode} {
        add $op_j$ to $opcodes$ \;
    }
    \If{$op_j$ reads data from memory which was written by an assignment opcode $op_k$} {
        add $op_k$ to $opcodes$ \;
        add $f_d(tr, op_k)$ to $opcodes$\;
    }
    \If{$op_j$ reads data from storage which was written by an assignment opcode $op_k$} {
        \If{$op_k$ is not visited} {
            add $op_k$ to $opcodes$\;
            \ForEach{trace $tr'$ contains $op_k$} {
                add $f_d(tr', op_k)$ to $opcodes$\;   
            }
        }
    }
}
return $opcodes$\;
\caption{$f_d(tr,op_i$)}
\label{algo:data_dependency_traverse}
\end{algorithm}

Identifying the set of opcodes which $op_i$ is data-dependent on is more complicated. Data dependency arises from 3 data sources, i.e., stack, memory and storage. In the following, we present our over-approximation based algorithm which traces data-flow on these data sources in order to capture data dependency. Although an opcode typically reads and writes data to the same data source, an opcode may write data to a different data source in some cases. That makes data-flow tracing complicated, i.e., data flows from stack to memory through \texttt{MSTORE}, memory to stack through \texttt{MLOAD}, stack to storage through \texttt{SSTORE} and storage to stack through \texttt{SLOAD}. Since only assignment opcodes (i.e., \texttt{SWAP}, \texttt{MSTORE}, and \texttt{SSTORE}) create data dependency, we thus design an algorithm to identify data-dependency based on the assignment opcodes in $tr$. 
The details are presented in the Algorithm \ref{algo:data_dependency_traverse}, which takes a symbolic trace $tr$ and opcode $op_i$ as input and returns a set of opcodes that $op_i$ is data-dependent on. 

Algorithm \ref{algo:data_dependency_traverse} systematically identifies those opcodes in $tr$ which taint $op_i$. An opcode $op_j$ is said to taint another opcode $op_i$ if $op_i$ reads data from stack indexes written by $op_j$, or there exists an opcode $op_t$ such that $op_j$ taints $op_t$ and $op_t$ taints $op_i$. For each $op_j$ that taints $op_i$, there are three possible dependency cases.
\begin{itemize}
	\item Stack dependency: $op_i$ is data-dependent on $op_j$ if $op_j$ is an assignment opcode (i.e., \texttt{SWAP}) (lines 3-4)
	\item Memory dependency: $op_j$ is data-dependent on $op_k$ if $op_j$ reads data from memory which was written by the assignment opcode $op_k$ (i.e., \texttt{MSTORE}) (lines 5-7)
	\item Storage dependency: $op_j$ is data-dependent on $op_k$ if $op_j$ reads data from storage which was written by the assignment opcode $op_k$ (i.e., \texttt{SSTORE}) (lines 8-12)
\end{itemize}
Note that the algorithm is recursive, i.e., if $op_k$ is added into the set of opcodes to be returned, a recursive call is made to further identify those opcode that $op_k$ is data-dependent on (lines 7 and 12). Further note that since storage is globally accessible, the analysis may be cross different traces in $Tr$ (line 11).

Algorithm \ref{algo:data_dependency_traverse} in general over-approximates. For instance, because memory and storage addresses are likely symbolic values, a reading address and a writing address are often incomparable, in which case we conservatively assume that the addresses may be the same. 
In other words, $R(op_j) \cap W(op_k) \neq \emptyset$ is true if either $R(op_j)$ or $W(op_k)$ is a symbolic address. 

\subsection{Fixing the Smart Contract}
Once the dependencies are identified, we check whether each symbolic $tr$ suffers from any of the vulnerabilities defined in Section~\ref{sec:probdef} and then fix the smart contract accordingly. In general, a smart contract is fixed as follows. Given a vulnerable trace $tr$, according to our definitions in Section~\ref{sec:probdef}, there must be an external call $op_c \in C$ in $tr$. Furthermore, there must be some other opcode $op$ that $op_c$ depends on which together makes $tr$ vulnerable (e.g., if $op$ is \texttt{SSTORE}, $tr$ suffers from reentrancy vulnerability; if $op$ is \texttt{ADD}, \texttt{SUB}, \texttt{MUL} or \texttt{DIV}, $tr$ suffers from arithmetic vulnerability). The idea is to introduce runtime checks right before $op$ so as to prevent the vulnerability. According to the type of vulnerability, the runtime checks are injected as follows.
\begin{itemize}
    \item To prevent intra-function reentrancy vulnerability, we add a modifier \texttt{nonReentrant} to the function $F$ containing $op$. Note that the \texttt{nonReentrant} modifier works as a mutex which blocks an attacker from re-entering $F$. To prevent cross-function reentrancy vulnerability, we add the modifier \texttt{nonReentrant} to the function containing $op$. The details of the fixing algorithm are presented in Algorithm~\ref{algo:fix_reentrancy} which takes a vulnerable trace $tr$ and the dependency relation $dp$ as inputs.
    \item To fix dangerous tx.origin vulnerability, we replace $op$ (i.e., \texttt{ORIGIN}) with \texttt{msg.sender} which returns address of the immediate account that invokes the function.
    \item To fix arithmetic vulnerability, we replace $op$ (i.e., \texttt{ADD}, \texttt{SUB}, \texttt{MUL}, \texttt{DIV}, or \texttt{EXP}) with a call to a safe math function which checks for overflow/underflow before performing the arithmetic operation. 
\end{itemize}
Note that in the case of reentrancy vulnerability and arithmetic vulnerability, if a runtime check fails (e.g., \texttt{assert(x > y)} which is introduced before \texttt{x - y} fails), the transaction reverts immediately and thus the vulnerability is prevent, although the gas spent on executing the transaction so far would be wasted. Further note while Algorithm~\ref{algo:fix_reentrancy} is applied to every vulnerable trace, the same fix (e.g., introducing \texttt{nonReentrant} on the same function) is applied once. We refer the readers to Section~\ref{sec:patchexample} for examples on how smart contracts are fixed.

The following establishes the soundness of our approach. 
\begin{theorem}
A smart contract fixed by Algorithm~\ref{algo:sguard} is free of intra-function reentrancy vulnerability, cross-function reentrancy vulnerability, dangerous tx.origin vulnerability, and arithmetic vulnerability. \hfill \qedsymbol
\end{theorem}
The proof of the theorem is sketched as follows. According to the Lemma~\ref{lem:Tr_is_complete}, given a smart contract $S$, if there are vulnerable traces, at least one of them is identified by \tool. Given how \tool{} fixes each kind of vulnerability, fixing all vulnerable traces in $Tr$ implies that all vulnerable traces are fixed in $S$.

We acknowledge that our approach does not achieve the preciseness as discussed in Section~\ref{sec:probdef}. That is, a trace which is not vulnerable may be affected by the fixes if it shares some opcodes with the vulnerable traces. For instance, an arithmetic opcode which is shared by a vulnerable trace and a non-vulnerable trace may be replaced with a safe version that checks for overflow. The non-vulnerable trace would revert in the case of an overflow even though the overflow might be benign. Such in-preciseness is an overhead to pay for security in our setting, along with the time and gas overhead. In Section~\ref{results}, we empirically evaluate that the overhead and show that they are negligible.

\begin{algorithm}[t]
\small
    let $tr \gets \langle s_0, op_0,\cdots, s_n, op_n, s_{n+1} \rangle$\;
    \ForEach{$i$ in $0..n$}{
        \If{$op_i \in C$}{
            \ForEach{$j$ in $i+1..n$} {
                \If{$op_j$ is \texttt{SSTORE} and $op_i$ depends on $op_j$ according to $dp$} {
                    \tcc{Fix intra-function reentrancy}
                    add modifier \texttt{nonReentrant} to the function containing $op_i$\;
                    \tcc{Fix cross-function reentrancy}
                    \ForEach{$op_s$ that $op_i$ depends on according to $dp$}{
                        \If{$op_s$ is \texttt{SSTORE}} {
                            add modifier \texttt{nonReentrant} to the function containing $op_s$\;
                        }
                    }
                }
            }
        }
    }
\caption{$fixReentrancy(tr, dp)$}
\label{algo:fix_reentrancy}
\end{algorithm}

\section{Implementation and Evaluation}
\label{results}
In this section, we present implementation details of \tool{} and then evaluate it with multiple experiments. 

\subsection{Implementation}
\label{sect:discussion}
\tool{} is implemented with around 3K lines of Node.js code. It is publicly available at~\href{https://github.com/reentrancy/sGuard}{GitHub}\footnote{\url{https://github.com/reentrancy/sGuard}}. It uses a locally installed compiler to compile a user-provided contract into a JSON file containing the bytecode, source-map and abstract syntax tree~(AST). The bytecode is used for detecting vulnerability, whereas the source-map and AST are used for fixing the smart contract at the source code level. In general, a source-map links an opcode to a statement and a statement to a node in an AST. Given a node in an AST, \tool{} then has the complete control on how to fix the smart contract. 



In addition to what is discussed in previous sections, the actual implementation of \tool{} has to deal with multiple complications. First, Solidity allows developers to interleave their codes with inline-assembly (i.e., a language using EVM machine opcodes). This allows fine-grained controls, as well as opens a door for hard-to-discover vulnerabilities (e.g., arithmetic vulnerabilities). We have considered fixing vulnerabilities with \tool{} (which is possible with efforts). However, it is not trivial for a developer to evaluate the correctness of our fixes as \tool{} would introduce opcodes into the inline-assembly. We do believe that any modification of the source code should be transparent to the users, and thus decide not to support fixing vulnerabilities inside inline-assembly. 

Second, \tool{} employs multiple heuristics to avoid useless fixes. For instance, given an arithmetic expression whose operands are concrete values (which may be the case of the expression is independent of user-inputs), \tool{} would not replace it with a function from safe math even if it is a part of a vulnerable trace. Furthermore, since the number of iterations to be unfolded for each loop depends on the number of assignment statements inside the loop, \tool{} identifies a number of cases where certain assignments can be safely ignored without sacrificing the soundness of our method. In particular, although we count \texttt{SSTORE}, \texttt{MSTORE} or \texttt{SWAP} as assignment statements in general, they are not in the following exceptional cases.
\begin{itemize}
    \item A \texttt{SWAP} is not counted if it is not mapped to an assignment statement according source-map;
    \item An assignment statement is not counted if its right-hand-side expression is a constant;
    \item An assignment statement is not counted if its left-hand-side expression is a storage variable (since dependency due to the storage variables is analyzed regardless of execution order).
\end{itemize}
In addition, \tool{} implements a strategy to estimate the value of memory pointers. A memory variable is always placed at a free memory pointer and it is never freed. However, the free pointer is often a symbolic value. That increases the complexity. To simplify the problem without missing dependency, \tool{} estimates the value of the free pointer $ptr$ if it is originally a symbolic value. That is, if the memory size of a variable is only known at run-time, we assume that it occupies 10 memory slots. The free pointer is calculated as $ptr_{n + 1} = 10 \times \texttt{0x20} + ptr_n$ where $ptr_n$ is the previous free pointer. If memory overlap occurs due to this assumption, additional dependencies are introduced, which may introduce false alarms, but never false negatives.

Lastly, \tool{} allows user to provide additional guide to generate contract-specific fixes. For instance, users are allowed to declare certain variables are critical variables so that it will be protected even if there is no dependency between the variable and external calls. 


\subsection{Evaluation}
In the following, we evaluate \tool{} through multiple experiments to answer the following research questions (RQ). Our test subjects include 5000 contracts whose verified source code are collected from EtherScan~\cite{etherscan}. This includes all the contracts after we filter 5000 incompilable contracts which contain invalid syntax or are implemented based on previous versions of Solidity (e.g., version 0.3.x). 
We systematically apply \tool{} to each contract. The timeout is set to be 5 minutes for each contract. Our experiments are conducted on with 10 concurrent processes and takes 6 hours to complete. All experiment results reported below are obtained on an Ubuntu 16.04.6 LTS machine with Intel(R) Core(TM) i9-9900 CPU @ 3.10GHz and 64GB of memory. \bigbreak

\begin{figure}[t]
    \centering
    \includegraphics[width=0.5\textwidth]{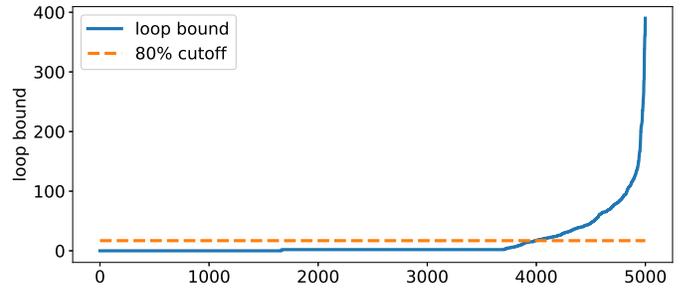}
    \caption{Loop bounds computed by \tool{}}
    \label{fig:path_explosion}
\end{figure}
\noindent \emph{RQ1: How bad is the path explosion problem?} 
Out of the 5000 contract, \tool~times out on 1767 (i.e., 35.34\%) contracts and successfully finish analyzing and fixing the remaining contracts within the time limit. Among them, 1590 contracts are deemed safe (i.e., they do not contain any external calls) and no fix is applied. The remaining 1643 contracts are fixed in one way or another. We note that 38 of the fixed contracts are incompilable. There are two reasons. First, the contract source-map may refer to invalid code locations if the corresponding smart contract has special characters (e.g., copyright and heart emoji). This turns out to be a bug of the Solidity compiler and has been reported. Second, the formats of AST across many solidity versions are slightly different, e.g., version 0.6.3 declares a function which is implemented with attribute \textit{implemented} while the attribute is absent in version 0.4.26. 
Note that the compiler version declared by \textit{pragma} keyword is not supported in the experiment setup  as \tool{} uses a list of compilers provided by solc-select~\cite{solc-select}. In the end, we experiment with 1605 smart contracts and report the findings.

Recall that the number of paths explored largely depend on the loop bounds. To understand why \tool~times out on 35.34\% of the contracts, we record the maximum loop bound for each of the 5000 smart contracts. 
Figure \ref{fig:path_explosion} summarizes the distribution of the loop bounds. 
From the figure, we observe that for 80\% of the contracts, the loop bounds are no more than 17. The loop bounds of the remaining 20\% contracts however vary quite a lot, e.g., with a maximum of 390. The average loop bound is 15, which shows that the default bounds in existing symbolic execution engines could be indeed insufficient. 
\bigbreak

\noindent \emph{RQ2: Is \tool~capable of pinpointing where fixes should be applied?} This question asks whether \tool{} is capable of precisely identifying where the fixes should be applied. Recall that \tool{} determines where to apply the fix based on the result of the dependency analysis, i.e., a precise dependency analysis would automatically imply that the fix will be applied at the right place. Furthermore, control dependency is straightforward and thus the answer to this question relies on the preciseness of the data dependency analysis. Data dependency analysis in Algorithm~\ref{algo:data_dependency_traverse} may introduce impreciseness (i.e., over-approximation) at lines 5 and 8 when checking the intersection of reading/writing addresses.
In \tool, the checking is implemented by transforming each symbolic address to a range of concrete addresses using the base address and the maximum offset. The over-approximation is only applied if at least one symbolic address is failed to transform due to nonstandard access patterns. If both symbolic addresses are successfully transformed, we can use the ranges of concrete addresses to precisely check the intersection and there is no over-approximation.
Thus, we can measure the over-approximation of our analysis by reporting the number of failed and successful address transformations.

\begin{figure}[t]
    \centering
    \includegraphics[width=0.5\textwidth]{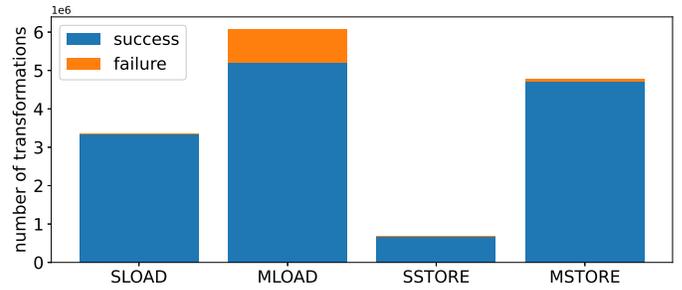}
    \caption{Memory and storage address transformations}
    \label{fig:tainting}
\end{figure}

Figure \ref{fig:tainting} summarizes our experiment results where each bar represents the number of failed and successful address transformations regarding the memory (i.e., \texttt{MLOAD}, \texttt{MSTORE}) and storage (i.e., \texttt{SLOAD}, \texttt{SSTORE}) opcodes. From the results, we observe that the percentage of successful transformations are 99.99\%, 85.58\%, 99.98\%, and 98.43\% for \texttt{SLOAD}, \texttt{MLOAD}, \texttt{SSTORE}, and \texttt{MSTORE} respectively. \texttt{MLOAD} has the worst accuracy among the four opcodes. This is mainly because some opcodes (e.g., \texttt{CALL}, and \texttt{CALLCODE}) may load different sizes of data on the memory. In this case, the \texttt{MLOAD} may depend on multiple \texttt{MSTORE}s, and it becomes even harder considering the size of loaded data is a symbolic value. Therefore, we simplify the analysis by returning true (hence over-approximates) if the size of loaded data is not \texttt{0x20}, a memory allocation unit size.
\bigbreak


\noindent \emph{RQ3: What is the runtime overhead of \tool's fixes?} This question is designed to measure the runtime overhead of \tool's fixes. Note that runtime overhead is often considered as a determining factor on whether to adopt additional checks at runtime. For instance, the C programming language has been refusing to introduce runtime overflow checks due to concerns on the runtime overhead, although many argue that it would reduce a significant number of vulnerabilities. The same question must thus be asked about \tool. Furthermore, runtime checks in smart contracts introduce not only time overhead but also gas overhead, i.e., gas must be paid for every additional check that is executed. Considering the huge number of transactions (e.g., 1.2 million daily transactions are reported on the Ethereum network~\cite{transaction-per-day}), each additional check may potential translate to large financial burden.  

To answer the question, we measure additional gas and computational time that users pay to deploy and execute the fixed contract in comparison with the original contract. That is, we download transactions from the Ethereum network and replicate them on our local network, and compare the gas/time consumption of the transactions. Among the 1605 smart contracts, 23 contracts are not considered as they are created internally. In the end, we replicate 6762 transactions of 1582 fixed contracts. We limit the number of transactions for each contract to a maximum of 10 such that the results are not biased towards those active contracts that have a huge number of transactions. 

Since our local setup is unable to completely simulate the actual Ethereum network (e.g., the block number and timestamps are different), a replicated transaction thus may end up being a revert. In our experiments, 3548 (52.47\%) transactions execute successfully and thus we report the results based on them. A close investigation shows that the remaining transactions fail due to the difference between our private network and the Ethereum network except 1 transaction, which fails because the size of the bytecode of the fixed contract exceeds the size limit~\cite{codesizelimit}.

\begin{figure}[t]
    \centering
    \includegraphics[width=0.5\textwidth]{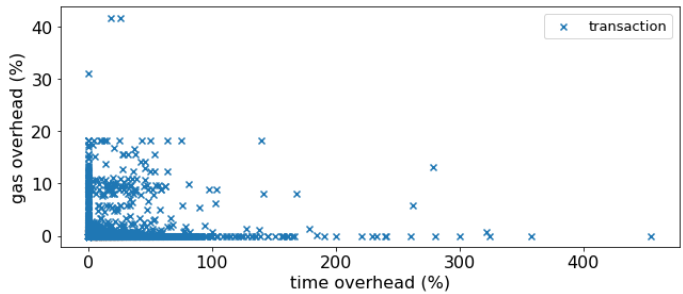}
    \caption{Overhead of fixed contracts}
    \label{fig:overhead}
\end{figure}


Figure~\ref{fig:overhead} summarizes our results. The x-axis and y-axis show the time overhead and gas overhead of each transaction respectively. The data shows that the highest gas overhead is 42\% while the lowest gas overhead is 0\%. On average, users have to pay extra 0.79\% gas to execute a transaction on the fixed contract. The highest and lowest time overhead are 455\% and 0\% respectively. On average, users have to wait extra 14.79\% time on a transaction. Based on the result, we believe that the overhead of fixing smart contracts using \tool{} is manageable, considering its security guarantee. 

For arithmetic vulnerabilities, there is a simplistic fix, i.e., add a check to every arithmetic operation. To see the difference between \tool~and such an approach, we conduct an additional experiment on the set of smart contracts that we successfully fixed (i.e., 1605 of them). We record the total number of bound checks added to the 4 arithmetic instructions (i.e., \texttt{ADD}, \texttt{SUB}, \texttt{MUL} and \texttt{DIV}) by \tool~and the simplistic approach. The results are shown in Table \ref{tab:bound_checks}, where column BC shows the number for the simplistic approach. 
\begin{table}[t]
    \footnotesize
    \centering
    \begin{tabular}{|p{1.2cm}|p{2cm}|p{2cm}|p{2cm}|}
      \hline
      \textbf{Instruction} & \textbf{\tool} & \textbf{BC} & \textbf{BC/\tool} \\ \hline
      \texttt{ADD} & 576 & 2245 & $3.9\times$\\ \hline
      \texttt{SUB} & 394 & 2125 & $5.39\times$\\ \hline
      \texttt{MUL} & 198 & 1423 & $7.19\times$\\ \hline
      \texttt{DIV} & 179 & 1508 & $8.42\times$\\ \hline
    \end{tabular}
    \caption{Total number of bound checks}
    \label{tab:bound_checks}
\end{table}
We observe that on average \tool~introduces $5.42$ times less bound checks than the simplistic approach. Since each bound check costs gas and time when executing a transaction, we consider such a reduction to be welcomed. 

\bigbreak
\begin{figure}[t]
    \centering
    \includegraphics[width=0.5\textwidth]{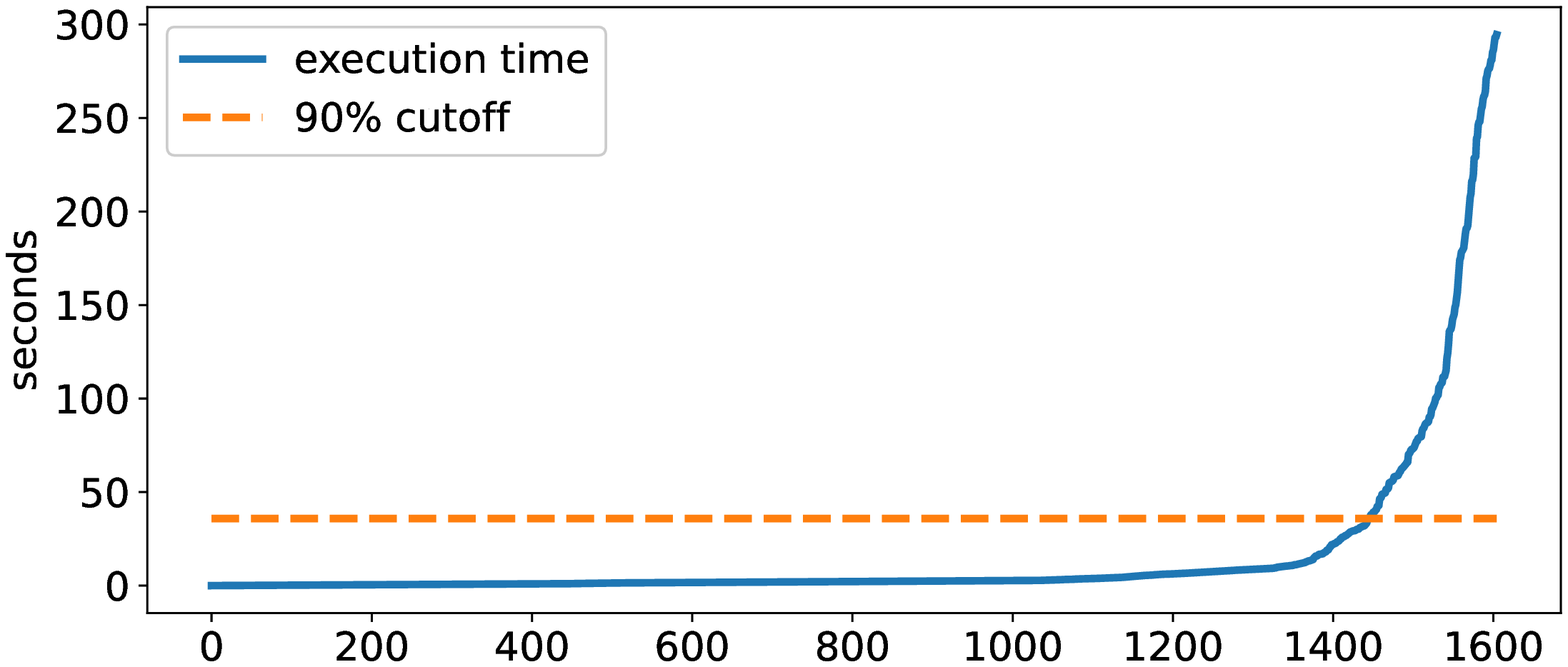}
    \caption{\tool~execution time}
    \label{fig:duration}
\end{figure}
\noindent \emph{RQ4: How long does \tool~take to fix a smart contract?} This question asks about the efficiency of \tool~itself. We measure the execution time of \tool~by recording time spending to fix each smart contract. Naturally, a more complicated contract (e.g., with more symbolic traces) takes more time to fix. Thus, we show how execution time varies for different contracts. Figure~\ref{fig:duration} summarizes our results, where each bar represents 10\% of smart contracts and y-axis shows the execution time in seconds. The contracts are sorted according to the execution time. From the figure, we observe that 90\% of contracts are fixed within 36 seconds. Among the different steps of \tool, \tool~spends most of the time to identify dependency (70.57\%) and find vulnerabilities (20.08\%). On average, \tool~takes 15 seconds to analyze and fix a contract.


\bigbreak
\noindent \emph{Manual inspection of results} To check the quality of the fix, we run an additional experiment on the top 10 \texttt{ERC20} tokens in the market. That is, we apply \tool{} to analyze and fix the contracts and then manually inspect the results to check whether the fixed contracts contain any of the vulnerabilities, i.e., whether \tool{} fails to prevent certain vulnerability or whether \tool{} introduce unnecessary runtime checks (which translates to considerable overhead given the huge number of transactions on these contracts).  The results are reported in Table~\ref{tab:high_profile} where column RE (respectively AE and TX) shows whether any reentrancy (respectively arithmetic, and tx.origin) vulnerability is discovered and fixed respectively; and the symbol \checkmark~and \xmark~denote yes and no respectively. The last column shows the number of symoblic traces explored.
\begin{table}[t]
    \footnotesize
    \centering
    \begin{tabular}{|p{0.75cm}|p{1cm}|p{0.75cm}|p{0.75cm}|p{0.75cm}|p{2cm}|}
      \hline
      \textbf{No.} & \textbf{Name} & \textbf{\#RE} & \textbf{\#AR} & \textbf{\#TX} & \textbf{Symbolic traces} \\ \hline
      \#1 & USDT & \xmark & \xmark & \xmark & 265 \\ \hline
      \#2 & LINK & \xmark & \xmark & \xmark & 291 \\ \hline
      \#3 & BNB & \xmark & \xmark & \xmark & 128  \\ \hline
      \#4 & HT & \xmark & \xmark & \xmark & 0  \\ \hline
      \#5 & BAT & \xmark & \checkmark & \xmark & 128  \\ \hline
      \#6 & CRO & \xmark & \xmark & \xmark & 401 \\ \hline
      \#7 & LEND & \xmark & \xmark & \xmark & 281 \\ \hline
      \#8 & KNC & \xmark & \xmark & \xmark & 443 \\ \hline
      \#9 & ZRX & \xmark & \xmark & \xmark & 0 \\ \hline
      \#10 & DAI & \xmark & \xmark & \xmark & 0 \\ \hline
    \end{tabular}
    \caption{Fixing results on the high profile contracts}
    \label{tab:high_profile}
\end{table}

We observe that the number of symbolic traces explored for three tokens HT, ZRX and DAI are 0. It is because these contracts contain no external calls and thus \tool{} stops immediately after scanning the bytecode. 
Among the remaining 7 tokens, six of them (i.e., LINK, BNB, CRO, LEND, KNC, and USDT) are found to be safe and thus no modification is made. 
One arithmetic vulnerability in the smart contracts BAT is reported and fixed by \tool{}. We confirm that a runtime check is added to prevent the discovered vulnerability. A close investigation however reveals that this vulnerability is unexploitable although it confirms to our definition. This is because the contract already has runtime checks. We further measure the overhead of the fix by executing 10 transactions obtained from the Ethereum network on the smart contract. The result shows that \tool~introduces a gas overhead of 18\%. Lastly, our manual investigation confirms that all of the contracts are free of the vulnerabilities.

\section{Related Work}
\label{related}
To the best of our knowledge, \tool{} is the first tool that aims to repair smart contracts in a provably correct way.

\tool{}~is closely related to the many works on automated program repair, which we highlight a few most relevant ones in the following. GenProg~\cite{Weimer09icse} applies evolutionary algorithm to search for program repairs. A candidate repair patch is considered successful if the repaired program passes all test cases in the test suite. In~\cite{kim2013automatic}, Dongsun \emph{et al.} presented PAR, which improves GenProg by learning fix patterns from existing human-written patches to avoid nonsense patches. In~\cite{abadi2009control}, Abadi \emph{et al.} automatically rewrites binary code to enforce control flow integrity (CFI). In~\cite{perkins2009automatically}, Jeff \emph{et al.} presented ClearView, which learns invariants from normal behavior of the application, generates patches and observes the execution of patched applications to choose the best patch. While there are many other program repair works, none of them focus on fixing smart contracts in a provably correct way.  

\tool{} is closely related to the many work on applying static analysis techniques on smart contracts. Securify~\cite{tsankov2018securify} and Ethainter~\cite{brent2020ethainter} are approaches which leverage a rewriting system (i.e., Datalog) to identify vulnerabilities through pattern matching. In terms of symoblic execution, in~\cite{luu2016making}, Luu \emph{et al.} presented the first engine to find potential security bugs in smart contracts. In~\cite{krupp2018teether}, Krupp and Rossow presented teEther which finds vulnerabilities in smart contracts by focusing on financial transactions. In~\cite{nikolic2018finding}, Nikolic \emph{et al.} presented MAI-AN, which focus on identifying trace-based vulnerabilities through a form of symoblic execution. In~\cite{torres2018osiris}, Torres \emph{et al.} presented Osiris which focuses on discovering integer bugs. Unlike these engines, \tool~not only detects vulnerabilities, but also fixes them automatically.

\tool~is related to some work on verifying and analyzing smart contracts. Zeus~\cite{kalra2018zeus} is a framework which verifies the correctness and fairness of smart contracts based on LLVM. Bhargavan \emph{et al.} proposed a framework to verify smart contracts by transforming the source code and the bytecode to an intermediate language called F*~\cite{bhargavan2016formal}. In~\cite{hirai2016formal}, the author used Isabelle/HOL to verify the Deed contract. In~\cite{frowis2017code}, the authors showed that only 40\% of smart contracts are trustworthy based on their call graph analysis. In~\cite{chen2017under}, Chen \emph{et al.} showed that most of the contracts suffer from some gas-cost programming patterns.

Finally, \tool~is remotely related to approaches on testing smart contracts. ContractFuzzer~\cite{Jiang:2018:CFS:3238147.3238177} is a fuzzing engine which checks 7 different types of vulnerabilities. sFuzz~\cite{nguyen2020sfuzz} is another fuzzer which extends ContractFuzzer by using feedback from test cases execution to generate new test cases.
\section{Conclusion}
\label{conclusion}
In this work, we propose an approach to fix smart contracts so that they are free of 4 kinds of common vulnerabilities. Our approach uses run-time information and is proved to be sound. The experiment results show the usefulness of our approach, i.e., \tool{} is capable of fixing contracts correctly while introducing only minor overhead. In the future, we intend to improve the performance of \tool~further with optimization techniques.

\bibliographystyle{IEEEtran}
\bibliography{ref}

\begin{thebibliography}{10}
\providecommand{\url}[1]{#1}
\csname url@samestyle\endcsname
\providecommand{\newblock}{\relax}
\providecommand{\bibinfo}[2]{#2}
\providecommand{\BIBentrySTDinterwordspacing}{\spaceskip=0pt\relax}
\providecommand{\BIBentryALTinterwordstretchfactor}{4}
\providecommand{\BIBentryALTinterwordspacing}{\spaceskip=\fontdimen2\font plus
\BIBentryALTinterwordstretchfactor\fontdimen3\font minus
  \fontdimen4\font\relax}
\providecommand{\BIBforeignlanguage}[2]{{%
\expandafter\ifx\csname l@#1\endcsname\relax
\typeout{** WARNING: IEEEtran.bst: No hyphenation pattern has been}%
\typeout{** loaded for the language `#1'. Using the pattern for}%
\typeout{** the default language instead.}%
\else
\language=\csname l@#1\endcsname
\fi
#2}}
\providecommand{\BIBdecl}{\relax}
\BIBdecl

\bibitem{Weimer09icse}
W.~Weimer, T.~Nguyen, C.~{Le Goues}, and S.~Forrest, ``Automatically finding
  patches using genetic programming,'' in \emph{Proceedings of the 31st
  International Conference on Software Engineering}, ser. ICSE '09, 2009, pp.
  364--374.

\bibitem{kim2013automatic}
D.~Kim, J.~Nam, J.~Song, and S.~Kim, ``Automatic patch generation learned from
  human-written patches,'' in \emph{2013 35th International Conference on
  Software Engineering (ICSE)}.\hskip 1em plus 0.5em minus 0.4em\relax IEEE,
  2013, pp. 802--811.

\bibitem{marginean2019sapfix}
A.~Marginean, J.~Bader, S.~Chandra, M.~Harman, Y.~Jia, K.~Mao, A.~Mols, and
  A.~Scott, ``Sapfix: Automated end-to-end repair at scale,'' in \emph{2019
  IEEE/ACM 41st International Conference on Software Engineering: Software
  Engineering in Practice (ICSE-SEIP)}.\hskip 1em plus 0.5em minus 0.4em\relax
  IEEE, 2019, pp. 269--278.

\bibitem{tsankov2018securify}
P.~Tsankov, A.~Dan, D.~Drachsler-Cohen, A.~Gervais, F.~Buenzli, and M.~Vechev,
  ``Securify: Practical security analysis of smart contracts,'' in
  \emph{Proceedings of the 2018 ACM SIGSAC Conference on Computer and
  Communications Security}, 2018, pp. 67--82.

\bibitem{brent2020ethainter}
L.~Brent, N.~Grech, S.~Lagouvardos, B.~Scholz, and Y.~Smaragdakis, ``Ethainter:
  a smart contract security analyzer for composite vulnerabilities.'' in
  \emph{PLDI}, 2020, pp. 454--469.

\bibitem{wood2014ethereum}
G.~Wood \emph{et~al.}, ``Ethereum: A secure decentralised generalised
  transaction ledger,'' \emph{Ethereum project yellow paper}, vol. 151, no.
  2014, pp. 1--32, 2014.

\bibitem{rsk}
\BIBentryALTinterwordspacing
{RSK}. [Online]. Available: \url{https://www.rsk.co/}
\BIBentrySTDinterwordspacing

\bibitem{hyperledger}
\BIBentryALTinterwordspacing
{Hyperledger}. [Online]. Available: \url{https://www.hyperledger.org/}
\BIBentrySTDinterwordspacing

\bibitem{deops199}
\BIBentryALTinterwordspacing
{A Postmortem on the Parity Multi-Sig Library Self-Destruct}. [Online].
  Available:
  \url{https://www.parity.io/a-postmortem-on-the-parity-multi-sig-library-self-destruct/}
\BIBentrySTDinterwordspacing

\bibitem{contractbugs}
\BIBentryALTinterwordspacing
{Thinking About Smart Contract Security}. [Online]. Available:
  \url{https://blog.ethereum.org/2016/06/19/thinking-smart-contract-security/}
\BIBentrySTDinterwordspacing

\bibitem{openzeppelin}
\BIBentryALTinterwordspacing
{OpenZeppelin}. [Online]. Available:
  \url{https://github.com/OpenZeppelin/openzeppelin-contracts}
\BIBentrySTDinterwordspacing

\bibitem{9152785}
\BIBentryALTinterwordspacing
J.~Jiao, S.~Kan, S.~Lin, D.~Sanan, Y.~Liu, and J.~Sun, ``Semantic understanding
  of smart contracts: Executable operational semantics of solidity,'' in
  \emph{2020 IEEE Symposium on Security and Privacy (SP)}.\hskip 1em plus 0.5em
  minus 0.4em\relax Los Alamitos, CA, USA: IEEE Computer Society, may 2020, pp.
  1695--1712. [Online]. Available:
  \url{https://doi.ieeecomputersociety.org/10.1109/SP40000.2020.00066}
\BIBentrySTDinterwordspacing

\bibitem{rodler2018sereum}
M.~Rodler, W.~Li, G.~Karame, and L.~Davi, ``Sereum: Protecting existing smart
  contracts against re-entrancy attacks,'' in \emph{Proceedings of the Network
  and Distributed System Security Symposium ({NDSS'19})}, 2019.

\bibitem{known-attacks}
\BIBentryALTinterwordspacing
{Known Attacks}. [Online]. Available:
  \url{https://consensys.github.io/smart-contract-best-practices/known_attacks/}
\BIBentrySTDinterwordspacing

\bibitem{nguyen2020sfuzz}
T.~D. Nguyen, L.~H. Pham, J.~Sun, Y.~Lin, and Q.~T. Minh, ``sfuzz: An efficient
  adaptive fuzzer for solidity smart contracts,'' in \emph{Proceedings of the
  42nd International Conference on Software Engineering (ICSE)}, 2020, pp.
  778--788.

\bibitem{chang2019scompile}
J.~Chang, B.~Gao, H.~Xiao, J.~Sun, Y.~Cai, and Z.~Yang, ``scompile: Critical
  path identification and analysis for smart contracts,'' in
  \emph{International Conference on Formal Engineering Methods}.\hskip 1em plus
  0.5em minus 0.4em\relax Springer, 2019, pp. 286--304.

\bibitem{mossberg2019manticore}
M.~Mossberg, F.~Manzano, E.~Hennenfent, A.~Groce, G.~Grieco, J.~Feist,
  T.~Brunson, and A.~Dinaburg, ``Manticore: A user-friendly symbolic execution
  framework for binaries and smart contracts,'' in \emph{2019 34th IEEE/ACM
  International Conference on Automated Software Engineering (ASE)}.\hskip 1em
  plus 0.5em minus 0.4em\relax IEEE, 2019, pp. 1186--1189.

\bibitem{dominator}
\BIBentryALTinterwordspacing
A worklist algorithm for dominators. [Online]. Available:
  \url{http://pages.cs.wisc.edu/~fischer/cs701.f08/lectures/Lecture19.4up.pdf}
\BIBentrySTDinterwordspacing

\bibitem{etherscan}
\BIBentryALTinterwordspacing
{Etherscan}. [Online]. Available: \url{https://etherscan.io/}
\BIBentrySTDinterwordspacing

\bibitem{solc-select}
\BIBentryALTinterwordspacing
{Solc-Select}. [Online]. Available: \url{https://github.com/crytic/solc-select}
\BIBentrySTDinterwordspacing

\bibitem{transaction-per-day}
\BIBentryALTinterwordspacing
{Ethereum transactions per day}. [Online]. Available:
  \url{https://etherscan.io/chart/tx}
\BIBentrySTDinterwordspacing

\bibitem{codesizelimit}
\BIBentryALTinterwordspacing
{EIP-170}. [Online]. Available:
  \url{https://github.com/ethereum/EIPs/blob/master/EIPS/eip-170.md}
\BIBentrySTDinterwordspacing

\bibitem{abadi2009control}
M.~Abadi, M.~Budiu, {\'U}.~Erlingsson, and J.~Ligatti, ``Control-flow integrity
  principles, implementations, and applications,'' \emph{ACM Transactions on
  Information and System Security (TISSEC)}, vol.~13, no.~1, pp. 1--40, 2009.

\bibitem{perkins2009automatically}
J.~H. Perkins, S.~Kim, S.~Larsen, S.~Amarasinghe, J.~Bachrach, M.~Carbin,
  C.~Pacheco, F.~Sherwood, S.~Sidiroglou, G.~Sullivan \emph{et~al.},
  ``Automatically patching errors in deployed software,'' in \emph{Proceedings
  of the ACM SIGOPS 22nd symposium on Operating systems principles}, 2009, pp.
  87--102.

\bibitem{luu2016making}
L.~Luu, D.-H. Chu, H.~Olickel, P.~Saxena, and A.~Hobor, ``Making smart
  contracts smarter,'' in \emph{Proceedings of the 2016 ACM SIGSAC Conference
  on Computer and Communications Security}.\hskip 1em plus 0.5em minus
  0.4em\relax ACM, 2016, pp. 254--269.

\bibitem{krupp2018teether}
J.~Krupp and C.~Rossow, ``teether: Gnawing at ethereum to automatically exploit
  smart contracts,'' in \emph{27th $\{$USENIX$\}$ Security Symposium
  ($\{$USENIX$\}$ Security 18)}, 2018, pp. 1317--1333.

\bibitem{nikolic2018finding}
I.~Nikoli{\'c}, A.~Kolluri, I.~Sergey, P.~Saxena, and A.~Hobor, ``Finding the
  greedy, prodigal, and suicidal contracts at scale,'' in \emph{Proceedings of
  the 34th Annual Computer Security Applications Conference}.\hskip 1em plus
  0.5em minus 0.4em\relax ACM, 2018, pp. 653--663.

\bibitem{torres2018osiris}
C.~F. Torres, J.~Sch{\"u}tte \emph{et~al.}, ``Osiris: Hunting for integer bugs
  in ethereum smart contracts,'' in \emph{Proceedings of the 34th Annual
  Computer Security Applications Conference}.\hskip 1em plus 0.5em minus
  0.4em\relax ACM, 2018, pp. 664--676.

\bibitem{kalra2018zeus}
S.~Kalra, S.~Goel, M.~Dhawan, and S.~Sharma, ``Zeus: Analyzing safety of smart
  contracts,'' in \emph{25th Annual Network and Distributed System Security
  Symposium (NDSS'18)}, 2018.

\bibitem{bhargavan2016formal}
K.~Bhargavan, A.~Delignat-Lavaud, C.~Fournet, A.~Gollamudi, G.~Gonthier,
  N.~Kobeissi, N.~Kulatova, A.~Rastogi, T.~Sibut-Pinote, N.~Swamy
  \emph{et~al.}, ``Formal verification of smart contracts: Short paper,'' in
  \emph{Proceedings of the 2016 ACM Workshop on Programming Languages and
  Analysis for Security}.\hskip 1em plus 0.5em minus 0.4em\relax ACM, 2016, pp.
  91--96.

\bibitem{hirai2016formal}
Y.~Hirai, ``Formal verification of deed contract in ethereum name service,''
  \emph{November-2016.[Online]. Available: https://yoichihirai. com/deed. pdf},
  2016.

\bibitem{frowis2017code}
M.~Fr{\"o}wis and R.~B{\"o}hme, ``In code we trust?'' in \emph{Data Privacy
  Management, Cryptocurrencies and Blockchain Technology}.\hskip 1em plus 0.5em
  minus 0.4em\relax Springer, 2017, pp. 357--372.

\bibitem{chen2017under}
T.~Chen, X.~Li, X.~Luo, and X.~Zhang, ``Under-optimized smart contracts devour
  your money,'' in \emph{2017 IEEE 24th International Conference on Software
  Analysis, Evolution and Reengineering (SANER)}.\hskip 1em plus 0.5em minus
  0.4em\relax IEEE, 2017, pp. 442--446.

\bibitem{Jiang:2018:CFS:3238147.3238177}
\BIBentryALTinterwordspacing
B.~Jiang, Y.~Liu, and W.~K. Chan, ``Contractfuzzer: Fuzzing smart contracts for
  vulnerability detection,'' in \emph{Proceedings of the 33rd ACM/IEEE
  International Conference on Automated Software Engineering}, ser. ASE
  2018.\hskip 1em plus 0.5em minus 0.4em\relax New York, NY, USA: ACM, 2018,
  pp. 259--269. [Online]. Available:
  \url{http://doi.acm.org/10.1145/3238147.3238177}
\BIBentrySTDinterwordspacing

\end{thebibliography}
\end{document}